\documentclass[pdflatex,sn-nature]{sn-jnl}

\usepackage{graphicx}%
\usepackage{multirow}%
\usepackage{amsmath,amssymb,amsfonts}%
\usepackage{amsthm}%
\usepackage{mathrsfs}%
\usepackage[title]{appendix}%
\usepackage{xcolor}%
\usepackage{textcomp}%
\usepackage{manyfoot}%
\usepackage{booktabs}%
\usepackage{algorithm}%
\usepackage{algorithmicx}%
\usepackage{algpseudocode}%
\usepackage{listings}%
\usepackage{txfonts}

\usepackage{bm}

\usepackage{adjustbox}

\theoremstyle{thmstyletwo}%

\theoremstyle{thmstylethree}%

\begin{document}

\newgeometry{textwidth=6.5in, outermargin=1in, innermargin=1in, top=1in, bottom=1in}

\footskip=20pt

\title{Hydrodynamic origins of symmetric swimming strategies}

\author*[1,2,3]{Takahiro Kanazawa}\email{kanazawa@uchicago.edu}

\author*[4]{Kenta Ishimoto}\email{kenta.ishimoto@math.kyoto-u.ac.jp}

\author*[1,2,5,6]{Kyogo Kawaguchi}\email{kyogo.kawaguchi@riken.jp}
\affil[1]{\small \orgdiv{Department of Physics}, \orgname{The University of Tokyo}, \orgaddress{\street{7-3-1 Hongo, Bunkyo-ku}, \city{Tokyo}, \postcode{113-0033}, \country{Japan}}}

\affil[2]{\small \orgdiv{Nonequilibrium Living Matter Laboratory}, \orgname{RIKEN Pioneering Research Institute}, \orgaddress{\street{2-1 Hirosawa}, \city{Wako}, \postcode{351-0198}, \country{Japan}}}

\affil[3]{\small \orgdiv{Department of Physics}, \orgname{The University of Chicago}, \orgaddress{\street{5720 South Ellis Avenue}, \city{Chicago}, \state{Illinois}, \postcode{60637}, \country{USA}}}

\affil[4]{\small \orgdiv{Department of Mathematics}, \orgname{Kyoto University}, \orgaddress{\street{Kitashirakawa-Oiwake-cho}, \city{Kyoto}, \postcode{606-8502}, \country{Japan}}}

\affil[5]{\small \orgdiv{Institute for Physics of Intelligence}, \orgname{The University of Tokyo}, \orgaddress{\street{7-3-1 Hongo, Bunkyo-ku}, \city{Tokyo}, \postcode{113-0033}, \country{Japan}}}

\affil[6]{\small \orgdiv{Universal Biology Institute}, \orgname{The University of Tokyo}, \orgaddress{\street{7-3-1 Hongo, Bunkyo-ku}, \city{Tokyo}, \postcode{113-0033}, \country{Japan}}}

\maketitle

\begin{abstract}

Efficient locomotion is important for the evolution of complex life~\cite{fritz2020evolution, brunet2023cell}, yet the physical principles selecting specific swimming strokes often remain entangled with biological constraints. In viscous fluids, the scallop theorem constrains the temporal organization of strokes~\cite{purcell1977life, lauga2011life}, but no analogous principle is known for their spatial structure, leaving the prevalence of symmetric gaits across diverse organisms~\cite{elgeti2015physics, lauga2009hydrodynamics} without a physical explanation. Here we show that spatial symmetry acts as an emergent organizing principle for efficiency in viscous fluids. By analysing deformable swimmers whose strokes are not constrained to any particular symmetry class~\cite{shapere1989geometry, shapere1989efficiencies}, we identify a hydrodynamic duality: symmetric and anti-symmetric strokes are dynamically equivalent, yielding identical speeds and efficiencies, which we prove are optimal among all strokes. We validate this using numerical simulations of Stokes flow~\cite{cortez2001method, cortez2005method3d}, demonstrating that these symmetry rules persist even in three-dimensional body plans. Our results suggest that the prevalence of symmetric and alternating gaits in nature reflects not merely a developmental constraint, but a physical optimality principle for locomotion in viscous environments, complementing developmental and neural constraints.
\end{abstract}

\section*{Introduction}

The evolution of animal motility marked a pivotal transition in the history of life, driven by the emergence of bilaterian body plans that coordinated left and right sides for directional movement \cite{fritz2020evolution, brunet2023cell, hollo2012manoeuvrability}. From the simplest flagellates to complex multicellular organisms, swimming strategies often display spatial symmetries \cite{elgeti2015physics, lara2026forces}, respecting their symmetric shape with respect to the swimming direction. While these patterns are often attributed to developmental constraints or neural architecture, it remains an open question whether they represent physically optimal strategies for interacting with a fluid environment, or simply convenient biological accidents.

In the microscopic realm, where viscosity dominates inertia, this question is particularly relevant. Physics dictates that propulsion at low Reynolds number requires time-irreversible (non-reciprocal) deformations \cite{purcell1977life}, a principle that underlies both theoretical descriptions of microswimming and the design of artificial swimmers \cite{shapere1989geometry, lauga2009hydrodynamics, ishimoto2012coordinate}. The scallop theorem constrains the temporal organization of swimming strokes: reciprocal motions produce no net displacement. Yet there is no analogous principle governing the spatial organization of strokes. Among the vast space of time-irreversible deformations, it remains unknown whether certain spatial symmetries are hydrodynamically preferred, or whether the bilateral coordination so prevalent in nature is incidental to the physics of propulsion.

This uncertainty persists because the geometric view of locomotion, where swimming arises from closed loops in the space of shapes \cite{shapere1989geometry, kelly1995geometric}, has largely been applied to restricted classes of motion. Most theoretical studies assume axisymmetric geometries or mirror-symmetric strokes that generate straight swimming by construction \cite{lighthill1952squirming, blake1971spherical, bae2010swimming, arroyo2012reverse}. While this geometric framework is well established for symmetric swimmers, it has not been applied to the full space of deformations without \emph{a priori} symmetry constraints. As a result, we lack a general understanding of how translational and rotational motion are organized when no symmetry is imposed on the swimming stroke. Specifically, it is unknown whether the alternating left--right motions prominent in nature cannot achieve higher efficiency than mirror-symmetric strokes due to the coupling of translation and rotation, or whether they represent an equally viable hydrodynamic strategy.

\begin{figure*}[t!]
\centering
\includegraphics[width=1.0\linewidth]{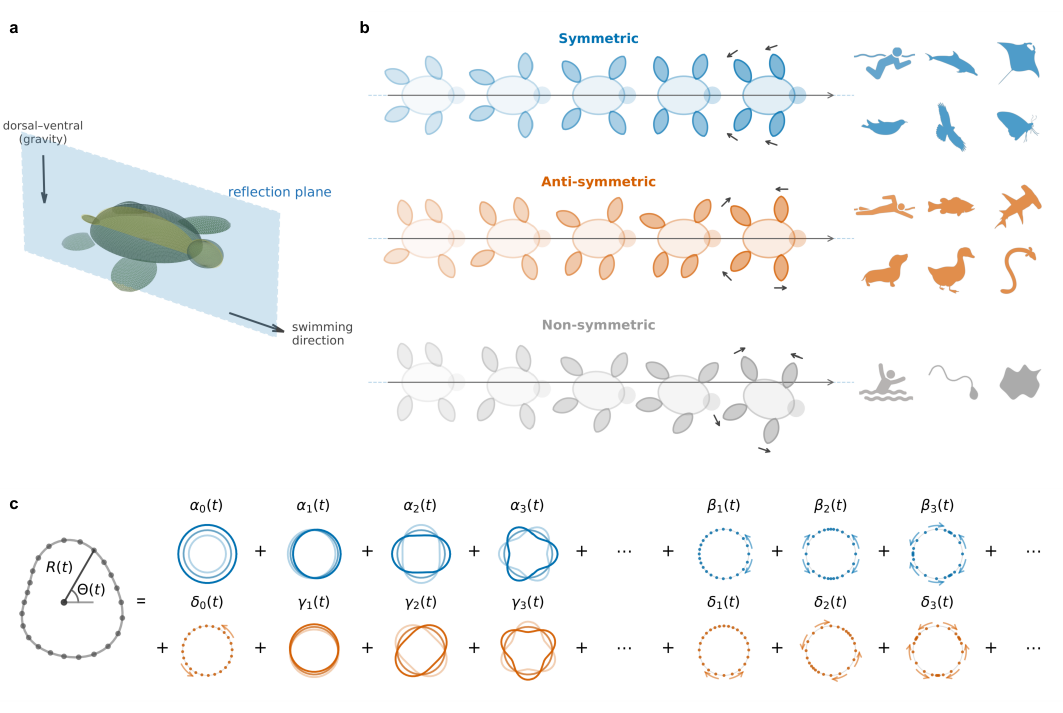}
\vspace{-1em}
\caption{Swimming strokes and their symmetry decomposition.
(a)~A bilaterally symmetric swimmer, with the sagittal (reflection) plane defined by the swimming direction and the dorsal--ventral axis.
(b)~Three classes of swimming strokes illustrated by montages of the body shape over one cycle, with representative animal silhouettes: symmetric strokes, anti-symmetric strokes, and non-symmetric strokes.
(c)~Fourier decomposition of the swimmer boundary. The radial displacement $R(\theta,t)$ and tangential coordinate $\Theta(\theta,t)$ are expanded in cosine and sine harmonics, with symmetric (even) modes ($\alpha_n$, $\beta_n$; blue) and anti-symmetric (odd) modes ($\gamma_n$, $\delta_n$; orange).
}
\label{fig1}
\end{figure*}

Here we address this question by developing a general framework for deformable swimming that treats translation and rotation on equal footing. By relaxing all symmetry constraints, we identify three generic classes of swimming strokes in the case of two dimensions: symmetric, anti-symmetric, and generic non-symmetric (Fig.~\ref{fig1}a,b). We show that symmetric and anti-symmetric strokes form a hydrodynamic dual pair: exchanging symmetric deformation modes for their anti-symmetric counterparts leaves the mean swimming speed and efficiency unchanged at leading order, despite distinct surface kinematics. By contrast, the optimal efficiency is attained by symmetric (or equivalently anti-symmetric) strokes, as generic non-symmetric strokes generate rotational drift that must be actively suppressed at an energetic cost. We validate these findings using numerical simulations of Stokes flow, demonstrating that this symmetry-based organization persists beyond the small-deformation regime and extends to three-dimensional swimmers. Together, our results identify symmetry not as a constraint, but as an emergent organizing principle for efficient locomotion in viscous environments.

\section*{General swimmers in two dimensions and hydrodynamic duality}

Many swimming organisms display bilateral symmetry not only in their body shape but also in their locomotor kinematics. The sagittal plane, defined by the swimming direction and the dorsal–ventral axis, acts as a mirror plane that organizes the deformations of the body surface (Fig.~\ref{fig1}a). This spatial organization naturally divides swimming strokes into three classes (Fig.~\ref{fig1}b). In \emph{symmetric strokes}, surface deformations are mirror-symmetric about the swimming axis at every instant, as in dolphins or the human breaststroke and butterfly (Fig.~\ref{fig1}b, top). In \emph{anti-symmetric strokes}, the left and right sides undergo deformation in opposite directions, producing undulatory motion as in fish or the human front crawl (Fig.~\ref{fig1}b, middle). Strokes that combine both symmetries are termed \emph{non-symmetric}, as seen in circling sperm and swimming amoebae (Fig.~\ref{fig1}b, bottom). More generally, the instantaneous boundary velocity of any stroke can be decomposed into components that are even or odd under reflection about the reflection plane: a stroke is symmetric if only even components are present at every instant, anti-symmetric if only odd components are present, and non-symmetric otherwise.

In two dimensions the reflection reduces to $y \to -y$ about the swimming axis $x$. To make this classification precise we consider deformable swimmers whose boundary undergoes small periodic deformations about a reference circle of radius $a$. The instantaneous configuration is described in polar coordinates (Fig.~\ref{fig1}c):
\begin{align}
R(\theta,t) &= a \left[ 1 + \sum_{n=0}^N \left( \alpha_n(t)\cos n\theta + \gamma_n(t)\sin n\theta \right) \right], \\
\Theta(\theta,t) &= \theta + \sum_{n=0}^N \left( \beta_n(t)\sin n\theta + \delta_n(t)\cos n\theta \right),
\end{align}
where $\theta$ is a material (Lagrangian) label identifying each point on the boundary, $R(\theta,t)$ is the radial distance encoding normal deformations, and $\Theta(\theta,t)$ is the angular position encoding tangential redistribution of material points along the surface. This representation provides a unified description of shape change and surface motion that accommodates arbitrary periodic deformations, including shapes with appendages and fins, provided the boundary does not self-intersect. Moreover, the polar decomposition naturally encodes the symmetry classification introduced above: the $\alpha_n$ and $\beta_n$ modes are even under $y\to-y$, while the $\gamma_n$ and $\delta_n$ modes are odd. Symmetric strokes activate only the even modes ${\alpha_n,\beta_n}$, anti-symmetric strokes only the odd modes ${\gamma_n,\delta_n}$, and non-symmetric strokes mix both sets.

To derive analytical results, we consider the swimming at low Reynolds number, where fluid motion is governed by the Stokes equations and propulsion arises entirely from cyclic shape changes. In this regime, swimming can be viewed as a geometric process: net motion results from closed loops in the space of shapes, with hydrodynamic interactions determining how deformations couple to translation and rotation \cite{shapere1989geometry}.
For a periodic swimming stroke of period $\tau$, the net translation and rotation accumulated over one cycle can be computed perturbatively in the deformation amplitudes. To leading order, the mean translational and angular velocities take the general form
\begin{align}
\langle U_i \rangle &= \frac{1}{\tau
}\int_0^\tau
\sum_{n,\sigma,\sigma'} 
\mathcal{U}_{n}^{\sigma,\sigma',i}
\left [\sigma_n(t)\dot{\sigma}'_{n+1}(t) - \dot{\sigma}_n(t)\sigma'_{n+1}(t) \right]\mathrm{d}t ,
\label{eq:U_i_main} \\
\langle \Omega \rangle &= \frac{1}{\tau}\int_0^\tau 
\sum_{n,\sigma,\sigma'} 
\mathcal{W}_{n}^{\sigma,\sigma'}
\left [\sigma_n(t)\dot{\sigma}'_{n}(t) - \dot{\sigma}_n(t)\sigma'_{n}(t) \right]\mathrm{d}t ,
\label{eq:Omega_main}
\end{align}
where $\sigma,\sigma'\in\{\alpha,\beta,\gamma,\delta\}$ labels deformation modes, $U_i$ denotes the translational velocity ($i=x,y$), and $\Omega$ is the angular velocity. These expressions make explicit a general geometric property of Stokes flow: net translation arises from couplings between deformation modes of adjacent order $n$ and $n+1$, whereas rotation arises from couplings between modes of the same order $n$. 
The explicit forms of the coefficient tensors $\mathcal{U},\mathcal{W}$, as well as the explicit form of the energy dissipation rate $\langle P \rangle$, are calculated and provided in the Supplementary Information.

\begin{figure*}[t!]
\centering
\includegraphics[width=1.0\linewidth]{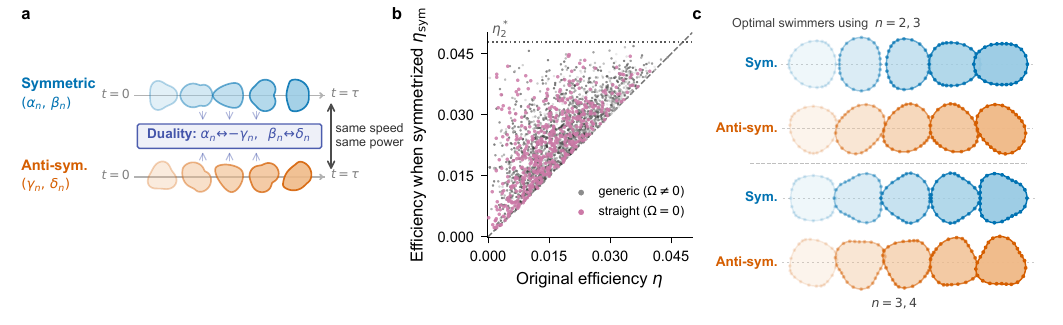}
\vspace{-1em}
\caption{Symmetry, hydrodynamic duality, and swimming efficiency.
(a) Schematic of the duality.
(b) Dimensionless efficiency $\eta=\langle U\rangle/\langle P\rangle$ comparing random swimmers using $n=2,3$ modes, with and without the constraint of $\langle \Omega\rangle=0$, with the symmetrized swimmers. The dotted line corresponds to the theoretical maximum efficiency $\eta^*_2$ under the constraint that the swimmer uses $n=2,3$ modes.
(c) Optimal straight swimmer within the symmetric (equivalently, anti-symmetric) subspace when restricted to the $n=2,3$ modes, or the $n=3,4$ modes, illustrating the corresponding shape cycle.}
\label{fig2}
\end{figure*}

Strikingly, this framework uncovers a \emph{hydrodynamic duality} connecting the two symmetry-preserving classes. We find that exchanging symmetric deformation modes for their anti-symmetric counterparts ($\alpha_n \leftrightarrow -\gamma_n; \beta_n \leftrightarrow \delta_n$) leaves the mean swimming speed and energetic cost unchanged at leading order, despite the entirely distinct surface kinematics. Symmetric and anti-symmetric swimming thus form a hydrodynamic dual pair: they represent dynamically equivalent strategies for a straight propulsion (Fig.~\ref{fig2}a). This invariance is a fundamental property of the hydrodynamic response; even for generic non-symmetric deformations, performing the duality transformation yields a partner stroke with identical speed and total dissipation. 

\section*{Energetic cost and efficiency of swimming}

To compare different swimming strategies, we define a dimensionless swimming efficiency $\eta = \langle U \rangle/\langle P \rangle$ as the ratio between the average forward motion produced during one cycle and the energy dissipated over that cycle~\cite{shapere1989efficiencies}. This measure captures the fundamental trade-off between propulsion and energetic cost in viscous environments, allowing different stroke patterns to be compared on equal footing. In the regime of small shape deformations considered here, both the swimming speed and the dissipated power scale similarly with the amplitude of the deformation, so their ratio remains well defined and provides a convenient measure of performance.

Efficient locomotion also requires suppression of rotational drift: rotation reorients the swimming direction without contributing to net translation, so only straight swimmers with $\langle \Omega \rangle = 0$ can achieve high efficiency. Generic non-symmetric strokes couple to rotation at the same order as translation, $O(\varepsilon^2)$, and suppressing this rotation requires additional deformation modes that dissipate energy without aiding propulsion. Symmetry-organized strokes avoid this competition by construction, making them natural candidates for optimal efficiency.

Intriguingly, this intuition can be formalized as a rigorous bound. We can prove that among all periodic strokes with vanishing mean rotation, no stroke can exceed the maximum swimming efficiency attained within the purely symmetric (or equivalently, purely anti-symmetric) subspace. Specifically, for any non-symmetric swimmer tuned to have zero net rotation, one can construct a corresponding pair of symmetric and anti-symmetric strokes whose efficiencies are at least as high (see Supplementary Information). Remarkably, the bound is tight: there exists a one-parameter family of non-symmetric swimmers that achieves the same efficiency as the symmetric stroke, by combining a symmetric and an anti-symmetric copy of the optimal stroke (see Supplementary Information). Generic non-symmetric strokes, however, do not satisfy this relation; we demonstrate numerically that randomly sampled non-symmetric strokes are consistently outperformed by their symmetry-organized counterparts (Fig.~\ref{fig2}b).

Having established that the optimal efficiency can be found within the symmetric (or anti-symmetric) subspace, we can then ask what deformation protocol maximizes efficiency. The optimization problem can be solved explicitly; in Fig.~\ref{fig2}c, we illustrate the motion of the optimal swimmer when restricting to the second and third, as well as third and fourth Fourier modes. Including higher-order modes further increases the efficiency, which converges to a limiting value $\eta^*_\infty  \approx 0.0563$ (Extended Data Fig.~\ref{figS1}). A similar limiting efficiency was first considered by Shapere and Wilczek for symmetric swimmers~\cite{shapere1989efficiencies}; we have now proved that there is a corresponding anti-symmetric swimmer that is equally optimal, and that generic non-symmetric strokes cannot surpass this efficiency.

\begin{figure*}[t!]
\centering
\includegraphics[width=1.0\linewidth]{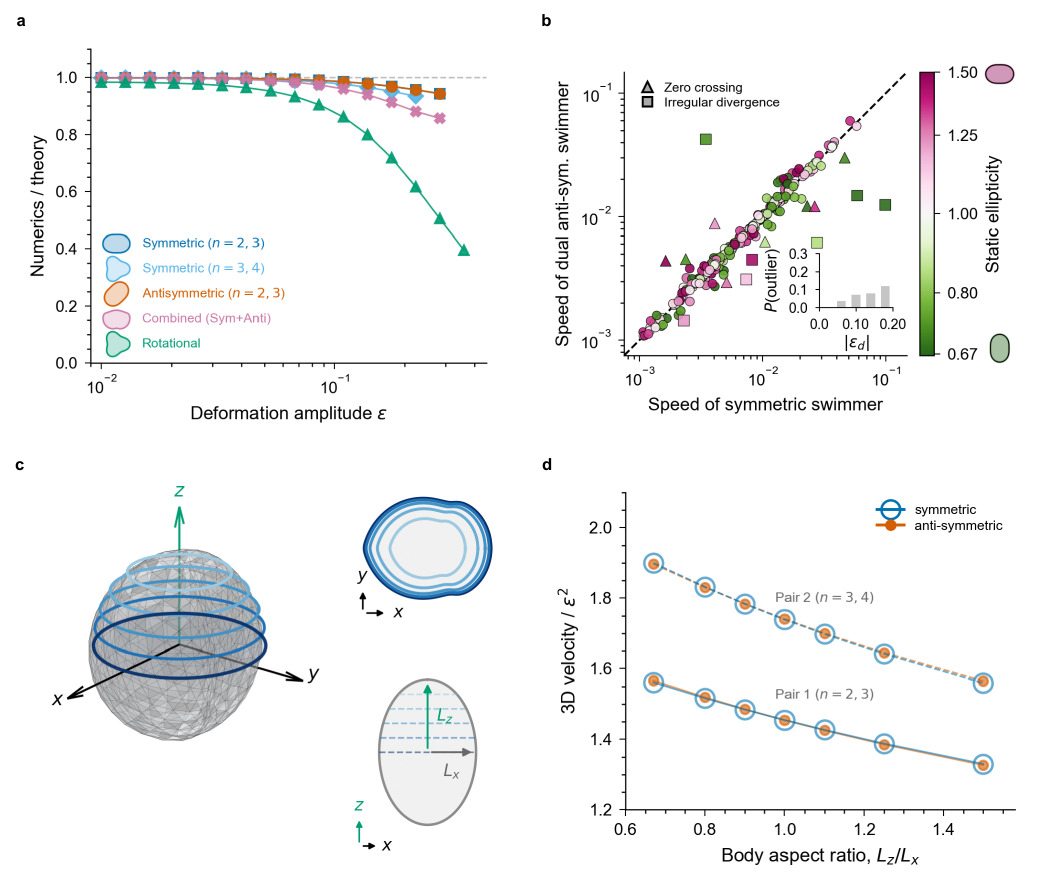}
\vspace{-1em}
\caption{Large deformations and extension to three-dimensional swimmers.
(a) Ratio of numerically computed to analytically predicted mean swimming speed ($\mathrm{Numerics}/\mathrm{theory}$) as a function of deformation amplitude $\varepsilon$, for five stroke classes. At large deformations the boundary parameterisation becomes non-monotonic and the shape approaches a singularity, making the swimming velocity ill-defined; this is why calculations for some strokes are truncated at $\varepsilon \sim 0.2$--$0.3$. 
(b) Swimming speed of the symmetric stroke versus that of its anti-symmetric dual for $256$ random samples of $M=576$ two-mode stroke pairs, coloured by the static ellipticity $e = (1+\varepsilon_d)/(1-\varepsilon_d)$ of the reference body, where $\varepsilon_d$ parameterises the undeformed shape as $R(\theta) = a(1 + \varepsilon_d \cos 2\theta)$. Deformation amplitude was fixed at $\varepsilon=0.15$. The unit of speed is $a/\tau$. Points on the dashed diagonal confirm equal speeds (hydrodynamic duality). Outliers deviating by more than a factor of $3/2$ are classified by their behaviour in an $\varepsilon$-sweep (Extended Data Fig.~\ref{figS4}): triangles denote near-cancellation of the leading-order velocity (zero crossing), squares denote irregular divergence of the numerics from the perturbative prediction. Inset: probability of such a deviation as a function of $|\varepsilon_d|$.    
(c) Schematic of the three-dimensional ``ellipsoid slice'' swimmer. Left: ellipsoidal body with the equatorial ($x$--$y$) swimming plane. Top right: cross-sections of the deformed body in the swimming ($x$--$y$) plane at several $z$-levels.
Bottom right: the $x$--$z$ cross-section showing the body silhouette and slice positions.
(d) Normalised swimming speed $|\langle U_x\rangle|/\varepsilon^2$ (in the unit of $L_x/\tau$) versus body aspect ratio $L_z/L_x$ for Pair~1 ($n=2,3$ modes, solid) and Pair~2 ($n=3,4$ modes, dashed), each in symmetric
(blue, open markers) and anti-symmetric (orange, filled markers) variants, at deformation
amplitude $\varepsilon=0.15$. 
}
\label{fig3}
\end{figure*}

\section*{Cases of large deformation and extension to three dimensions}

The analytical results so far are derived within a small-deformation expansion, raising the question of whether the symmetry-based organization they reveal persists beyond the perturbative regime. To address this, we perform numerical simulations using a boundary integral formulation of Stokes flow, which allows direct computation of translational and rotational motion for prescribed surface deformations. We imposed the regularization method introduced in \cite{cortez2001method, cortez2005method3d}, with the key hyperparameters tuned per stroke class by minimising a mesh-convergence indicator (Extended Data Fig.~\ref{figS2}). 

For sufficiently small deformation amplitudes, the numerical results quantitatively agree with the analytical predictions across all stroke classes (Fig.~\ref{fig3}a). The residual discrepancy between theory and numerics arises from spatial and temporal discretization and decreases systematically with increasing resolution, in a manner that depends on the stroke class and regularization parameter (Extended Data Fig.~\ref{figS3}; see also Supplementary Information). The qualitative structure imposed by symmetry remains robust at larger deformations (Fig.~\ref{fig3}b): symmetric and anti-symmetric strokes continue to exhibit closely matched swimming speeds even at $\varepsilon = 0.15$, with substantial differences appearing only when the swimmer's static shape deviates significantly from a disk (measured by the ellipticity shown in Fig.~\ref{fig3}b; see Extended Data Fig.~\ref{figS4} for a detailed $\varepsilon$-sweep analysis of individual outlier swimmers).

We identify two distinct mechanisms by which duality breaks down at large deformations (Fig.~\ref{fig3}b): (i)~near-cancellation of the leading-order velocity coefficient, causing higher-order terms to dominate the speed ratio, and (ii)~irregular divergence of the numerical velocity from the perturbative prediction, indicating rapid growth of nonlinear corrections for particular mode combinations (Extended Data Fig.~\ref{figS4}). Both mechanisms require deformation amplitudes $\varepsilon \gtrsim 0.10$--$0.15$ and are absent in the small-deformation regime where the perturbation theory applies.

To assess the generality of these findings, we next consider a class of three-dimensional swimmers motivated by real bilaterian body plans. In these swimmers, a dorsal--ventral axis is fixed, for example by gravity or environmental cues, while the dominant deformations occur within a transverse plane (Fig.~\ref{fig3}c). This construction preserves the essential symmetry structure of the two-dimensional theory while incorporating a third spatial dimension. Concretely, we model the swimmer as a prolate spheroid whose surface points are deformed by the same azimuthal Fourier modes as in 2D, applied identically at each latitude.

Numerical simulations of such three-dimensional swimmers show that the symmetry-based organization identified in two dimensions remains predictive. We tested two independent stroke pairs, Pair~1 ($n=2,3$ modes) and Pair~2 ($n=3,4$ modes), across body aspect ratios from oblate ($L_z/L_x = 0.67$) to prolate ($L_z/L_x = 1.50$). In all cases, symmetric and anti-symmetric strokes within each pair yield closely matched swimming speeds, agreeing to within $0.4\%$ across tested amplitudes (Fig.~\ref{fig3}d). While we do not have analytical proof of duality in three dimensions, our simulations indicate that hydrodynamic duality is robust under the natural extension to 3D.

\section*{Discussion and conclusion}

Beyond identifying optimal swimming strokes, our results clarify why certain classes of motion are generically selected in viscous environments. The sub-optimality of non-symmetric strokes is not merely quantitative but structural: breaking bilateral organization generically introduces rotation at leading order, while suppressing this rotation requires additional deformation modes that do not contribute to propulsion. Symmetry-organized strokes avoid this competition by construction, allowing propulsion to be achieved without the need for fine-tuned compensations. This provides a physical explanation for why efficient swimming strategies cluster within low-dimensional, symmetry-organized subspaces of the full space of possible deformations.

We consider that the symmetry-based perspective is particularly relevant to early stages of animal evolution, when controlled muscular motion first emerged \cite{hollo2012manoeuvrability, budd2017origin}. Early multicellular organisms may have possessed limited neural control, coarse actuation, and deformable bodies lacking rigid appendages. Under such conditions, locomotor strategies that were robust to imprecision and did not require precise cancellation of unwanted motion would have been strongly favoured. While chiral or helical swimming is widespread among unicellular microorganisms~\cite{shiba2008ca2,lauga2006swimming}, organisms that grow larger and use whole-body deformations for propulsion are subject to gravity, which provides a persistent dorsal--ventral axis favouring bilaterally organized strokes over chiral ones, precisely the scenario captured by our three-dimensional model (Fig.~\ref{fig3}c,d). Our results indicate that such bilaterally organized motions naturally achieve efficient propulsion while remaining insensitive to small variations in timing or amplitude, without the need for active suppression of rotation.

From this viewpoint, the rise of bilaterian body plans may have been facilitated not only by developmental or ecological advantages~\cite{finnerty2005bilateral}, but also by hydrodynamic and symmetry-based constraints on locomotion in viscous environments. Notably, symmetric coordination of motion is not restricted to bilaterians. Jellyfish propel themselves through radially symmetric bell contractions~\cite{costello2021hydrodynamics}, and ctenophores swim by coordinating rows of cilia that beat in phase along each comb row~\cite{tamm2014cilia}, or by bilaterian-like symmetric strokes~\cite{gemmell2019ctenophore}. In each case, the organism's locomotor strategy appears to respect or exploit the symmetry of its body plan rather than break it. Our framework, which identifies the physical advantage of symmetry-organized strokes in the simplest bilateral setting, may reflect a more general principle: that efficient self-propulsion in viscous fluids generically favors deformation protocols that are organized by the symmetry group of the swimmer.

At the same time, our work highlights a broader and relatively unexplored direction: symmetry classifications in locomotion have largely focused on body shape, whereas the symmetry of motion, that is, the spatiotemporal organization of deformations, remains comparatively understudied. For example, it is an interesting observation that nearly all aerial flyers (e.g., bats, birds, and insects) appear to be symmetric in their dynamics \cite{sane2003aerodynamics, dudley2000biomechanics}, 
whereas swimmers in water may be either symmetric or anti-symmetric. One possible explanation is that the negative effects of torque are more pronounced in air than in fluids \cite{dudley2000biomechanics}, though this remains outside the scope of our present analysis, which is restricted to the viscous regime. 

In three dimensions, the space of possible motion symmetries is considerably richer than the symmetric and anti-symmetric classes identified here, and may provide a natural framework for understanding efficient locomotion in non-bilaterian organisms as well. Exploring how higher-dimensional motion symmetry groups constrain propulsion and efficiency represents an important direction for future work.

\clearpage
\newpage

\backmatter



\section*{Methods}
\label{sec:methods}

\subsection*{Analytical framework}

We consider a two-dimensional deformable swimmer whose boundary undergoes small periodic deformations about a reference circle of radius $a$, parameterized in polar coordinates by radial and tangential displacements expanded in Fourier modes up to order $N$. The fluid is governed by the Stokes equations, solved via a stream function satisfying the biharmonic equation. Boundary conditions on the swimmer surface are imposed perturbatively in the deformation amplitude $\varepsilon$, yielding the instantaneous translational velocity $(U_x, U_y)$ and angular velocity $\Omega$ to $O(\varepsilon^2)$.

The time-averaged dynamics over one swimming period are computed using the geometric framework of Shapere and Wilczek~\cite{shapere1989geometry}, in which net displacement arises from the curvature (field strength) of a gauge connection on the space of shapes. This procedure accounts for the non-commutativity of translation and rotation in the Lie group SE(2). The resulting expressions for $\langle U_x \rangle$, $\langle U_y \rangle$, and $\langle \Omega \rangle$ are given in Eqs.~\eqref{eq:U_x_final}--\eqref{eq:Omega_final} of the Supplementary Information, from which the coefficient tensors in Eqs.~\eqref{eq:U_i_main} and \eqref{eq:Omega_main} follow directly.

The time-averaged power dissipation $\langle P \rangle$ is computed from the surface stress at leading order in $\varepsilon$, enabling a closed-form expression for the swimming efficiency $\eta = \langle U \rangle / \langle P \rangle$. The proof that symmetric strokes are at least as efficient as any non-symmetric stroke with $\langle \Omega \rangle = 0$ follows from decomposing an arbitrary stroke into symmetric and anti-symmetric components and applying the triangle inequality (see Supplementary Information). Optimal strokes within the symmetric subspace are obtained by solving an eigenvalue problem for the efficiency, yielding the maximum efficiency $\eta^*_n$ for each pair of adjacent Fourier modes $n$ and $n+1$, which converges to $\eta^*_\infty = \sqrt{2}/(8\pi) \approx 0.0563$ for large $n$. We solve the optimization problem explicitly for modes $n=2$ and $n=3$, obtaining closed-form expressions for the optimal deformation amplitudes (see Supplementary Information, Eqs.~\eqref{eq:eta_star}--\eqref{eq:evec_n3}).

\subsection*{Numerical simulations}

Swimming velocities for finite-amplitude deformations were computed using a boundary integral formulation of the Stokes equations with a regularized Stokeslet kernel~\cite{cortez2001method, cortez2005method3d}, in both two and three dimensions. The swimmer boundary was discretized into $M$ material points (2D) or $N$ quasi-uniform surface points generated by a Fibonacci sphere rule (3D). At each time step, the no-slip boundary condition and the force- and torque-free constraints yield a linear system for the unknown surface force densities and rigid-body velocities $(\mathbf{U}, \boldsymbol{\Omega})$, which was solved by direct linear algebra.

The regularization parameter $\varepsilon_{\mathrm{reg}}$, which sets the smoothing length $\delta = \varepsilon_{\mathrm{reg}} h$ relative to the local mesh spacing $h$, was determined empirically for each stroke class using a mesh-consistency criterion: we selected the value minimizing the fractional change in computed swimming speed between coarse ($M_c = 36$) and fine ($M_f = 72$) spatial resolutions, and verified stability at higher resolution ($M = 1728$). This per-stroke optimization of the regularization effectively suppresses the leading discretization error and enables quantitative comparison between numerical results and analytical predictions across all stroke classes (Extended Data Fig.~\ref{figS2},~\ref{figS3}).

For three-dimensional simulations, we employed an ``ellipsoid slice'' swimmer in which deformations are prescribed on each horizontal cross-section as planar Fourier-mode shape changes, while the dorsal-ventral extent is set by an ellipsoidal envelope with aspect ratio $L_z/L_x$. Simulations were performed for two independent stroke pairs (modes $n = 2,3$ and $n = 3,4$), each in symmetric and anti-symmetric variants, across aspect ratios $L_z/L_x \in [0.67, 1.50]$ with deformation amplitude $\varepsilon = 0.15$. Production runs used $N=480$ surface points and $T=64$ time steps per period. Regularization was tuned using coarse ($N_c=180$) and fine ($N_f=360$) resolutions. 

Full derivations, explicit coefficient expressions, and detailed convergence analyses are provided in the Supplementary Information.

\bmhead{Acknowledgments}

Pictograms and organism silhouettes in Fig.~\ref{fig1}b were obtained from Silhouette Illust (https://www.silhouette-illust.com) and Silhouette Design (https://kage-design.com), which provide free-to-use silhouette resources. 

K.I. acknowledges JSPS KAKENHI (Grant No. 24K21517) and the Japan Science and Technology Agency (JST), FOREST (Grant No. JPMJFR212N) and CREST (Grant No. JPMJCR25Q1).
This work was also supported by JSPS KAKENHI Grant Numbers JP19H05275, JP23H00095, and JP25H01361, and JST FOREST Program, Grant Number JPMJFR2435 to K.K.





\clearpage
\newpage

\section*{Supplementary Information}

\section{Analytical derivations of swimming velocity for small deformations}

The asymptotic swimming velocity for small deformations from a circle is calculated following Blake~\cite{blake1971self}.
Here we do not impose symmetry about the $x$ axis for generality.
Then the swimmer can in general move in the $x$ and $y$ directions and rotate.

\subsection{Equations of motion}

The fluid with zero Reynolds number is governed by the Stokes equation,
\begin{align}
    \nabla p &= \mu \nabla^2 \mathbf{u} ,
    \label{eq:Stokes} \\
    \nabla \cdot \mathbf{u} &= 0 ,
\end{align}
where $p$ is the pressure, $\mu$ is the viscosity of the fluid, and $\mathbf{u}$ is the fluid velocity.
In the current two-dimensional setup, we can introduce a stream function $\psi$ such that the biharmonic equation is satisfied,
\begin{equation}
    \nabla^4 \psi = 0 .
    \label{eq:biharmonic}
\end{equation}
The Laplacian $\nabla^2$ can be written in polar coordinates as
\begin{align}
    \nabla^2 &= \frac{\partial^2}{\partial x^2} + \frac{\partial^2}{\partial y^2} \\
    &= \frac{\partial^2}{\partial r^2} + \frac{1}{r} \frac{\partial}{\partial r} + \frac{1}{r^2} \frac{\partial^2}{\partial \theta^2}
\end{align}
where we have used the relation between Cartesian coordinates and polar coordinates
\begin{align}
    x &= r \cos{\theta} , \\
    y &= r \sin{\theta} .
\end{align}
The general solution of Eq.~\eqref{eq:biharmonic} is given as (e.g., \cite{wood2019general})
\begin{align}
    \psi (r,\theta) &= \tilde{a} + \tilde{b}r^2 + \tilde{c}\ln{r} + \tilde{d}r^2\ln{r} + \tilde{e}\theta + \tilde{f}r^2 \nonumber \\
    &\quad + \left( \frac{\tilde{g}}{r} + \tilde{h}r + \tilde{j}r^3 + \tilde{k}r\ln{r} + \tilde{l}r\theta \right) \cos{\theta} \nonumber \\
    &\quad + \left( \frac{\tilde{m}}{r} + \tilde{n}r + \tilde{o}r^3 + \tilde{p}r\ln{r} + \tilde{q}r\theta \right) \sin{\theta} \nonumber \\
    &\quad + \sum_{n=2}^\infty \left( \tilde{A}_n r^{-n} + \tilde{B}_n r^n + \tilde{C}_n r^{2-n} + \tilde{D}_n r^{2+n} \right) \cos{n \theta} \nonumber \\
    &\quad + \sum_{n=2}^\infty \left( \tilde{E}_n r^{-n} + \tilde{F}_n r^n + \tilde{G}_n r^{2-n} + \tilde{H}_n r^{2+n} \right) \sin{n \theta} ,
    \label{eq:stream_func_general}
\end{align}
where $\tilde{a}, \dots, \tilde{q}$ and $\{ \tilde{A}_n, \dots, \tilde{H}_n \}_n$ are constants.
It is easy to show that the velocity is related to the stream function as
\begin{align}
    u &= \frac{1}{r} \frac{\partial \psi}{\partial \theta} ,
    \label{eq:u_stream_func} \\
    v &= - \frac{\partial \psi}{\partial r} .
    \label{eq:v_stream_func}
\end{align}
Here, $u$ and $v$ are the radial and tangential components of the fluid velocity, respectively.
Notice that the governing equations are dependent only on the quantities of the interested time (i.e., quasi-static).

We denote the swimming velocity of the swimmer in the laboratory frame, where the fluid velocity vanishes infinitely far from the swimmer, as $\mathbf{U} = (U_x, U_y)^\mathrm{T}$.
Note that $U_x$ and $U_y$ are the $x$ and $y$ components of the velocity, where the directions (i.e., $x$ and $y$) are defined in the body-fixed frame of the swimmer.
The swimmer (or precisely the swimmer frame) in general rotates in the laboratory frame and we denote this angular velocity as $\Omega$.
Then from the boundary condition at infinity in the laboratory frame,
\begin{align}
    u \to 0, \ \ \ rv \to 0 \ \ \ (r \to \infty),
\end{align}
the relevant terms of Eq.~\eqref{eq:stream_func_general} can be rewritten as
\begin{equation}
    \psi = a_0 \theta + \frac{a_1}{r} \sin{\theta} + \sum_{n=2}^\infty \left( \frac{a_n}{r^n} + \frac{b_n}{r^{n-2}} \right) \sin{n \theta} + \frac{c_1}{r} \cos{\theta} + \sum_{n=2}^\infty \left( \frac{c_n}{r^n} + \frac{d_n}{r^{n-2}} \right) \cos{n \theta} ,
\end{equation}
where we have set the constant term as zero without loss of generality.
In the swimmer frame, on the other hand, we need to add terms to satisfy the boundary condition at infinity yielding
\begin{equation}
    \psi = - U_x r \sin{\theta} + U_y r \cos{\theta} + \frac{1}{2} \Omega r^2 + a_0 \theta + \frac{a_1}{r} \sin{\theta} + \sum_{n=2}^\infty \left( \frac{a_n}{r^n} + \frac{b_n}{r^{n-2}} \right) \sin{n \theta} + \frac{c_1}{r} \cos{\theta} + \sum_{n=2}^\infty \left( \frac{c_n}{r^n} + \frac{d_n}{r^{n-2}} \right) \cos{n \theta} .
    \label{eq:stream_func_swimmer}
\end{equation}
One can check that the stream function given in Eq.~\eqref{eq:stream_func_swimmer} satisfies the boundary condition at infinity by using Eqs.~\eqref{eq:u_stream_func} and \eqref{eq:v_stream_func}.

We assume that the boundary condition of the fluid velocity on the swimmer surface is given in the form of Fourier series as
\begin{align}
    u (a, \theta) &= \sum_{n=0}^\infty ( A_n \cos{n \theta} + C_n \sin{n \theta} ) ,
    \label{eq:u_boundary_cond} \\
    v (a, \theta) &= \sum_{n=0}^\infty ( B_n \sin{n \theta} + D_n \cos{n \theta})
    \label{eq:v_boundary_cond}
\end{align}
in a frame of reference attached to the swimmer (the swimmer frame).
We denote the radius of the undeformed swimmer as $a$.
Note that $B_0$ and $C_0$ can be set to zero without loss of generality.
In the next section, we derive an explicit expression of the coefficients $\{ A_n , B_n , C_n, D_n \}_n$ by means of small parameter expansions.

By imposing the boundary condition on the swimmer surface [Eqs.~\eqref{eq:u_boundary_cond} and \eqref{eq:v_boundary_cond}], we can determine unknown swimming (angular) velocities and coefficients in Eq.~\eqref{eq:stream_func_swimmer} as functions of $\{ A_n , B_n , C_n, D_n \}_n$ as
\begin{align}
    U_x &= \frac{1}{2} (B_1 - A_1) ,
    \label{eq:U_x_coeff} \\
    U_y &= -\frac{1}{2} (C_1 + D_1) ,
    \label{eq:U_y_coeff} \\
    \Omega &= -\frac{D_0}{a} ,
    \label{eq:Omega_coeff} \\
    a_0 &= aA_0 , \\
    a_1 &= \frac{a^2}{2} (A_1 + B_1) , \\
    a_n &= \frac{a^{n+1}}{2} \left(B_n - \frac{n-2}{n}A_n\right) \ \ (n \geq 2) , \\
    b_n &= \frac{a^{n-1}}{2} (A_n - B_n) \ \ (n \geq 2) , \\
    c_1 &= \frac{a^2}{2} (D_1 - C_1) , \\
    c_n &= \frac{a^{n+1}}{2} \left(D_n + \frac{n-2}{n} C_n\right) \ \ (n \geq 2) , \\
    d_n &= -\frac{a^{n-1}}{2} (C_n + D_n) \ \ (n \geq 2)
\end{align}
from Eqs.~\eqref{eq:u_stream_func} and \eqref{eq:v_stream_func} at $r=a$.

Therefore, we have determined the fluid velocity as
\begin{align}
    u(r, \theta) &= - U_x \cos{\theta} - U_y \sin{\theta} + A_0 \frac{a}{r} + \frac{1}{2} (A_1+B_1) \frac{a^2}{r^2} \cos{\theta} - \frac{1}{2} (D_1 - C_1) \frac{a^2}{r^2} \sin{\theta} \nonumber \\
    &\quad + \sum_{n=2}^\infty \frac{1}{2} A_n \cos{n\theta} \left[ - (n-2) \frac{a^{n+1}}{r^{n+1}} + n \frac{a^{n-1}}{r^{n-1}} \right] + \sum_{n=2}^\infty \frac{1}{2} B_n \cos{n\theta} \left[ n \frac{a^{n+1}}{r^{n+1}} - n \frac{a^{n-1}}{r^{n-1}} \right] \nonumber \\
    &\quad - \sum_{n=2}^\infty \frac{1}{2} C_n \sin{n\theta} \left[ (n-2) \frac{a^{n+1}}{r^{n+1}} - n \frac{a^{n-1}}{r^{n-1}} \right] - \sum_{n=2}^\infty \frac{1}{2} D_n \sin{n\theta} \left[ n \frac{a^{n+1}}{r^{n+1}} - n \frac{a^{n-1}}{r^{n-1}} \right] ,
    \label{eq:u_A-D}\\
    v(r, \theta) &= U_x \sin{\theta} - U_y \cos{\theta} - \Omega r + \frac{1}{2} (A_1+B_1) \frac{a^2}{r^2} \sin{\theta} + \frac{1}{2} (D_1-C_1) \frac{a^2}{r^2} \cos{\theta} \nonumber \\
    &\quad + \sum_{n=2}^\infty \frac{1}{2} A_n \sin{n\theta} \left[ - (n-2) \frac{a^{n+1}}{r^{n+1}} + (n-2) \frac{a^{n-1}}{r^{n-1}} \right] + \sum_{n=2}^\infty \frac{1}{2} B_n \sin{n\theta} \left[ n \frac{a^{n+1}}{r^{n+1}} - (n-2) \frac{a^{n-1}}{r^{n-1}} \right] \nonumber \\
    &\quad + \sum_{n=2}^\infty \frac{1}{2} C_n \cos{n\theta} \left[ (n-2) \frac{a^{n+1}}{r^{n+1}} - (n-2) \frac{a^{n-1}}{r^{n-1}} \right] + \sum_{n=2}^\infty \frac{1}{2} D_n \cos{n\theta} \left[ n \frac{a^{n+1}}{r^{n+1}} - (n-2) \frac{a^{n-1}}{r^{n-1}} \right]
    \label{eq:v_A-D}
\end{align}
with $U_x = (B_1-A_1)/2$, $U_y = -(C_1+D_1)/2$, and $\Omega = -D_0/a$.

\subsection{Surface deformations and instantaneous swimming dynamics}

As the Stokes equation is quasi-static, we derive an instantaneous dynamics of the swimmer in this section and will take its time average in the following section.
Thus, our goal in this section is to derive $U_x$, $U_y$, and $\Omega$ as functions of parameters characterizing the swimmer deformations.

Our swimmer has a circular shape with its radius $a$ without deformations.
We consider the general deformation of the swimmer with both radial and angular deformations up to the Fourier modes $n=N$ given as
\begin{align}
    R &= a \left[ 1 + \sum_{n=0}^N \left( \alpha_n(t) \cos{n\theta} + \gamma_n(t) \sin{n\theta} \right) \right] , \\
    \Theta &= \theta + \sum_{n=0}^N \left( \beta_n(t) \sin{n\theta} + \delta_n(t) \cos{n\theta} \right) ,
\end{align}
where $(R,\Theta)$ is the deformed position in polar coordinates.
We can regard $(a,\theta)$ as a (Lagrangian) label of the material points.
The Fourier coefficients $\{ \alpha_n(t), \beta_n(t), \gamma_n(t), \delta_n(t) \}_n$ characterize the deformations.
The normal (radial) deformation is characterized by $\alpha_n(t)$ and $\gamma_n(t)$, while the tangential (angular) deformation by $\beta_n(t)$ and $\delta_n(t)$.
As explained in the main text, $\alpha_n(t)$ and $\beta_n(t)$ [$\gamma_n(t)$ and $\delta_n(t)$] correspond to \emph{symmetric} (\emph{anti-symmetric}) strokes, respectively.
Note that $\beta_0(t)$ and $\gamma_0(t)$ can be set to zero without loss of generality.
From now on, we omit the explicit time dependence and simply write $\alpha_n$, $\beta_n$, $\gamma_n$, and $\delta_n$.
For explicit calculations of an asymptotic form, the deformations are considered to be small, that is, $\alpha_n, \beta_n, \gamma_n, \delta_n = O(\varepsilon)$, where $\varepsilon \ll 1$.

We aim to evaluate $U_x$, $U_y$, $\Omega$ [i.e., $D_0$, $A_1$, $B_1$, $C_1$, and $D_1$ from Eqs.~\eqref{eq:U_x_coeff}-\eqref{eq:Omega_coeff}] up to $O(\varepsilon^2)$.
Here we use a Taylor series expansion,
\begin{align}
    u(a,\theta) &= u(R,\Theta) - (R-a) \left. \frac{\partial u}{\partial r} \right|_{(a,\theta)} - (\Theta - \theta) \left. \frac{\partial u}{\partial \theta} \right|_{(a,\theta)} + O(\varepsilon^3) ,
    \label{eq:u_Taylor}  \\
    v(a,\theta) &= v(R,\Theta) - (R-a) \left. \frac{\partial v}{\partial r} \right|_{(a,\theta)} - (\Theta - \theta) \left. \frac{\partial v}{\partial \theta} \right|_{(a,\theta)} + O(\varepsilon^3)
    \label{eq:v_Taylor}
\end{align}
with
\begin{align}
    u(R,\Theta) &= \dot{R} = a \sum_{n=0}^N \left( \dot{\alpha}_n \cos{n\theta} + \dot{\gamma}_n \sin{n\theta} \right) , \\
    v(R,\Theta) &= R \dot{\Theta} = a \left[ 1 + \sum_{n=0}^N \left( \alpha_n \cos{n\theta} + \gamma_n \sin{n\theta} \right) \right] \sum_{n=0}^N \left( \dot{\beta}_n \sin{n\theta} + \dot{\delta}_n \cos{n\theta} \right) ,
\end{align}
where $\dot{f}$ denotes a time derivative of $f$, $\dot{f} = \partial f/\partial t$.
By comparing these with Eqs.~\eqref{eq:u_boundary_cond} and \eqref{eq:v_boundary_cond}, we have at the leading order, $O(\varepsilon)$,
\begin{align}
    A_n &= a \dot{\alpha}_n ,
    \label{eq:A_n_first} \\
    B_n &= a \dot{\beta}_n , \\
    C_n &= a \dot{\gamma}_n , \\
    D_n &= a \dot{\delta}_n
    \label{eq:D_n_first}
\end{align}
for $n=0, \dots, N$.
By carefully evaluating $\partial u/\partial r$, $\partial u/\partial \theta$, $\partial v/\partial r$, and $\partial v/\partial \theta$ in Eqs.~\eqref{eq:u_Taylor} and \eqref{eq:v_Taylor} from Eqs.~\eqref{eq:u_A-D} and \eqref{eq:v_A-D}, and using our first order solutions [i.e., Eqs.~\eqref{eq:A_n_first}-\eqref{eq:D_n_first}], we finally obtain on the order of $O(\varepsilon^2)$
\begin{align}
    \frac{D_0}{a} &= \sum_{n=1}^N \frac{1}{2} \left[ (n-2) \alpha_n \dot{\gamma}_n + 2n \alpha_n \dot{\delta}_n - (n-2) \gamma_n \dot{\alpha}_n + 2n \gamma_n \dot{\beta}_n + n \beta_n \dot{\delta}_n - n \delta_n \dot{\beta}_n \right] , \\
    \frac{A_1}{a} &= \alpha_0 \dot{\alpha}_1 + \alpha_0 \dot{\beta}_1 - \delta_0 \dot{\gamma}_1 + \alpha_1 \dot{\alpha}_0 \nonumber \\
    &\quad + \sum_{n=1}^{N-1} \frac{1}{2} \left[ \alpha_n \dot{\alpha}_{n+1} + (n+1) \alpha_n \dot{\beta}_{n+1} + \gamma_n \dot{\gamma}_{n+1} - (n+1) \gamma_n \dot{\delta}_{n+1} + (n+1) \beta_n \dot{\alpha}_{n+1} - (n+1) \delta_n \dot{\gamma}_{n+1} \right. \nonumber \\
    &\quad \qquad \qquad \left. + \alpha_{n+1} \dot{\alpha}_n + n \alpha_{n+1} \dot{\beta}_n + \gamma_{n+1} \dot{\gamma}_n - n \gamma_{n+1} \dot{\delta}_n + n \beta_{n+1} \dot{\alpha}_n - n \delta_{n+1} \dot{\gamma}_n \right] , \\
    \frac{B_1}{a} &= \alpha_0 \dot{\alpha}_1 + 2 \alpha_0 \dot{\beta}_1 + \delta_0 \dot{\delta}_1 \nonumber \\
    &\quad + \sum_{n=1}^{N-1} \frac{1}{2} \left[ - (n-1) \alpha_n \dot{\alpha}_{n+1} + 2(n+1) \alpha_n \dot{\beta}_{n+1} - (n-1) \gamma_n \dot{\gamma}_{n+1} - 2(n+1) \gamma_n \dot{\delta}_{n+1} + (n+1) \beta_n \dot{\beta}_{n+1} + (n+1) \delta_n \dot{\delta}_{n+1} \right. \nonumber \\
    &\quad \qquad \qquad \left. + (n-2) \alpha_{n+1} \dot{\alpha}_{n} - 2n \alpha_{n+1} \dot{\beta}_n + (n-2) \gamma_{n+1} \dot{\gamma}_n + 2n \gamma_{n+1} \dot{\delta}_n - n \beta_{n+1} \dot{\beta}_n - n \delta_{n+1} \dot{\delta}_n \right] , \\
    \frac{C_1}{a} &= \alpha_0 \dot{\gamma}_1 - \alpha_0 \dot{\delta}_1 + \delta_0 \dot{\alpha}_1 + \gamma_1 \dot{\alpha}_0 \nonumber \\
    &\quad + \sum_{n=1}^{N-1} \frac{1}{2} \left[ \alpha_n \dot{\gamma}_{n+1} - (n+1) \alpha_n \dot{\delta}_{n+1} - \gamma_n \dot{\alpha}_{n+1} - (n+1) \gamma_n \dot{\beta}_{n+1} + (n+1) \beta_n \dot{\gamma}_{n+1} + (n+1) \delta_n \dot{\alpha}_{n+1} \right. \nonumber \\
    &\quad \qquad \qquad \left. - \alpha_{n+1} \dot{\gamma}_n + n \alpha_{n+1} \dot{\delta}_n + \gamma_{n+1} \dot{\alpha}_n + n \gamma_{n+1} \dot{\beta}_n - n \beta_{n+1} \dot{\gamma}_n - n \delta_{n+1} \dot{\alpha}_n \right] , \\
    \frac{D_1}{a} &= - \alpha_0 \dot{\gamma}_1 + 2 \alpha_0 \dot{\delta}_1 - \delta_0 \dot{\beta}_1 \nonumber \\
    &\quad + \sum_{n=1}^{N-1} \frac{1}{2} \left[ (n-1) \alpha_n  \dot{\gamma}_{n+1} + 2(n+1) \alpha_n \dot{\delta}_{n+1} - (n-1) \gamma_n \dot{\alpha}_{n+1} + 2(n+1) \gamma_n \dot{\beta}_{n+1} + (n+1) \beta_n \dot{\delta}_{n+1} - (n+1) \delta_n \dot{\beta}_{n+1} \right. \nonumber \\
    &\quad \qquad \qquad \left. + (n-2) \alpha_{n+1} \dot{\gamma}_n + 2n \alpha_{n+1} \dot{\delta}_n - (n-2) \gamma_{n+1} \dot{\alpha}_n + 2n \gamma_{n+1} \dot{\beta}_n + n \beta_{n+1} \dot{\delta}_n - n \delta_{n+1} \dot{\beta}_n \right] .
\end{align}
We have used some trigonometric identities such as
\begin{align}
    \cos{n\theta} \cos{m\theta} &= \frac{1}{2} \left[ \cos{(n+m)\theta} + \cos{(n-m)\theta} \right] , \\
    \cos{n\theta} \sin{m\theta} &= \frac{1}{2} \left[ \sin{(n+m)\theta} - \sin{(n-m)\theta} \right] , \\
    \sin{n\theta} \cos{m\theta} &= \frac{1}{2} \left[ \sin{(n+m)\theta} + \sin{(n-m)\theta} \right] , \\
    \sin{n\theta} \sin{m\theta} &= \frac{1}{2} \left[ - \cos{(n+m)\theta} + \cos{(n-m)\theta} \right] .
\end{align}

We therefore obtain the swimmer velocities as
\begin{align}
    U_x &= \frac{1}{2} (B_1 - A_1) \\
    &= \frac{1}{2} a (\dot{\beta}_1 - \dot{\alpha}_1) + \frac{1}{2} a (\alpha_0 \dot{\beta}_1 + \delta_0 \dot{\delta}_1 + \delta_0 \dot{\gamma}_1 - \alpha_1 \dot{\alpha}_0) \nonumber \\
    &\quad + \sum_{n=1}^{N-1} \frac{1}{4} a \left[ - n \alpha_n \dot{\alpha}_{n+1} + (n+1) \alpha_n \dot{\beta}_{n+1} - n \gamma_n \dot{\gamma}_{n+1} - (n+1) \gamma_n \dot{\delta}_{n+1} \right. \nonumber \\
    &\quad \qquad \qquad + (n+1) \beta_n \dot{\beta}_{n+1} + (n+1) \delta_n \dot{\delta}_{n+1} - (n+1) \beta_n \dot{\alpha}_{n+1} + (n+1) \delta_n \dot{\gamma}_{n+1} \nonumber \\
    &\quad \qquad \qquad + (n-3) \alpha_{n+1} \dot{\alpha}_{n} - 3n \alpha_{n+1} \dot{\beta}_n + (n-3) \gamma_{n+1} \dot{\gamma}_n + 3n \gamma_{n+1} \dot{\delta}_n \nonumber \\
    &\quad \qquad \qquad \left. - n \beta_{n+1} \dot{\beta}_n - n \delta_{n+1} \dot{\delta}_n - n \beta_{n+1} \dot{\alpha}_n + n \delta_{n+1} \dot{\gamma}_n \right] + O(\varepsilon^3) ,
    \label{eq:U_x} \\
    U_y &= - \frac{1}{2} (C_1 + D_1) \\
    &= - \frac{1}{2} a (\dot{\gamma}_1 + \dot{\delta}_1) - \frac{1}{2} a (\alpha_0 \dot{\delta}_1 + \delta_0 \dot{\alpha}_1 + \gamma_1 \dot{\alpha}_0 - \delta_0 \dot{\beta}_1 ) \nonumber \\
    &\quad - \sum_{n=1}^{N-1} \frac{1}{4} a \left[ n \alpha_n \dot{\gamma}_{n+1} + (n+1) \alpha_n \dot{\delta}_{n+1} - n \gamma_n \dot{\alpha}_{n+1} + (n+1) \gamma_n \dot{\beta}_{n+1} \right. \nonumber \\
    &\quad \qquad \qquad + (n+1) \beta_n \dot{\gamma}_{n+1} + (n+1) \delta_n \dot{\alpha}_{n+1} + (n+1) \beta_n \dot{\delta}_{n+1} - (n+1) \delta_n \dot{\beta}_{n+1} \nonumber \\
    &\quad \qquad \qquad + (n-3) \alpha_{n+1} \dot{\gamma}_n + 3n \alpha_{n+1} \dot{\delta}_n - (n-3) \gamma_{n+1} \dot{\alpha}_n + 3n \gamma_{n+1} \dot{\beta}_n \nonumber \\
    &\quad \qquad \qquad \left. - n \beta_{n+1} \dot{\gamma}_n - n \delta_{n+1} \dot{\alpha}_n + n \beta_{n+1} \dot{\delta}_n - n \delta_{n+1} \dot{\beta}_n \right] + O(\varepsilon^3) ,
    \label{eq:U_y} \\
    \Omega &= - \frac{D_0}{a} \\
    &= - \dot{\delta}_0 - \sum_{n=1}^N \frac{1}{2} \left[ (n-2) \alpha_n \dot{\gamma}_n + 2n \alpha_n \dot{\delta}_n - (n-2) \gamma_n \dot{\alpha}_n + 2n \gamma_n \dot{\beta}_n + n \beta_n \dot{\delta}_n - n \delta_n \dot{\beta}_n \right] + O(\varepsilon^3).
    \label{eq:Omega}
\end{align}
Note that with $\gamma_n=\delta_n=0$, we recover the results in Blake~\cite{blake1971self}.

\subsection{Time-averaged swimming dynamics}

We here calculate the time-averaged dynamics for temporally periodic deformations.
In general, non-commutativity between the translation and the rotation yields nontrivial contributions to the time-averaged quantities~\cite{shapere1989geometry, shapere1989efficiencies, ishimoto2025bending}.

Following Ishimoto et al.~\cite{ishimoto2025bending}, we calculate the dynamics averaged over the period of the swimming gait.
We first introduce a $3 \times 3$ matrix representation of the Lie group $\mathcal{G} \in \mathrm{SE}(2)$ as
\begin{equation}
    \mathcal{G} =
    \begin{pmatrix}
    \mathbf{R}_\Theta & \mathbf{X} \\
    (0,0) & 1 \\
    \end{pmatrix}
\end{equation}
and the associated Lie algebra se(2).
Here $\mathbf{R}_\Theta$ is the two-dimensional rotation matrix by an angle $\Theta$ and $\mathbf{X}$ is the displacement.
The elements of se(2), $\hat{\mathcal{A}}$, act on SE(2) as
\begin{equation}
    \dot{\mathcal{G}} = \mathcal{G} \hat{\mathcal{A}} .
    \label{eq:diff_G_cal}
\end{equation}
Here, $\hat{\mathcal{A}}$ has its matrix representation,
\begin{equation}
    \hat{\mathcal{A}} =
    \begin{pmatrix}
    0 & \Omega & U_x \\
    -\Omega & 0 & U_y \\
    0 & 0 & 0
    \end{pmatrix}
    .
\end{equation}
From Eqs.~\eqref{eq:U_x}, \eqref{eq:U_y}, and \eqref{eq:Omega}, $\hat{\mathcal{A}}$ can be decomposed into a form of sum as
\begin{equation}
    \hat{\mathcal{A}} = \sum_\mu \hat{\mathcal{H}}_\mu \dot{\sigma}_\mu ,
\end{equation}
where $\sigma_\mu$ is the shape variable, that is, $\sigma_\mu \in \{ \alpha_n, \beta_n, \gamma_n, \delta_n \}_n$.
Here, a gauge field or Stokes connection, $\hat{\mathcal{H}}_\mu$, represented by a $3 \times 3$ matrix, depends only on $\{ \sigma_\mu \}_\mu$.
For a single period of gait, whose integral path in the shape space is denoted as $C$, the solution of the differential equation [Eq.~\eqref{eq:diff_G_cal}] is written as
\begin{align}
    \mathcal{G}(\tau) &= \mathcal{G}(0) \bar{\mathrm{P}} \exp{\left[ \int_0^\tau \hat{\mathcal{A}}(t) dt \right]} \\
    &= \mathcal{G}(0) \bar{\mathrm{P}} \exp{\left[ \sum_\mu \int_C \hat{\mathcal{H}}_\mu(\sigma) d\sigma_\mu \right]}
\end{align}
with $\bar{\mathrm{P}}$ being the reverse path ordering operator.
For small deformations, we have up to $O(\varepsilon^2)$
\begin{equation}
    \mathcal{G}(\tau) = \mathcal{G}(0) \exp{\left[ \sum_{\mu, \nu} \int_{S_C} \hat{\mathcal{F}}_{\mu \nu} (\sigma) d\sigma_\mu d\sigma_\nu \right]} + O(\varepsilon^3) .
\end{equation}
Here, $S_C$ represents the region confined by $C$ and the field strength or Stokes curvature, $\hat{\mathcal{F}}_{\mu \nu}$, is given by
\begin{equation}
    \hat{\mathcal{F}}_{\mu \nu} = -\frac{\partial \hat{\mathcal{H}}_\mu}{\partial \sigma_\nu} + \frac{\partial \hat{\mathcal{H}}_\nu}{\partial \sigma_\mu} - [\hat{\mathcal{H}}_\mu , \hat{\mathcal{H}}_\nu] ,
\end{equation}
where the bracket represents the matrix commutator, $[\hat{\mathcal{H}}_\mu , \hat{\mathcal{H}}_\nu] = \hat{\mathcal{H}}_\mu \hat{\mathcal{H}}_\nu - \hat{\mathcal{H}}_\nu \hat{\mathcal{H}}_\mu$.
Therefore, the averaged velocity is given by
\begin{equation}
    \langle \hat{\mathcal{A}} \rangle = \frac{1}{2} \sum_{\mu, \nu} \hat{\mathcal{F}}_{\mu \nu} \langle \sigma_\mu \dot{\sigma}_\nu \rangle ,
\end{equation}
where $\langle f \rangle$ denotes a time average of $f$,
\begin{equation}
    \langle f \rangle = \frac{1}{\tau} \int_0^\tau f(t) dt
\end{equation}
with $\tau$ being the period.

Following the procedure described above, we can explicitly calculate $\hat{\mathcal{A}}$, $\hat{\mathcal{H}}_\mu$, and $\hat{\mathcal{F}}_{\mu \nu}$, and thus $\langle U_x \rangle$, $\langle U_y \rangle$, and $\langle \Omega \rangle$.
Notice that $\hat{\mathcal{F}}_{\mu \nu} = - \hat{\mathcal{F}}_{\nu \mu}$ by definition and thus $\hat{\mathcal{F}}_{\mu \mu} = 0$.
In fact, the non-commutativity of the matrices $[\hat{\mathcal{H}}_\mu, \hat{\mathcal{H}}_\nu]$ appears only for the coupling between $\delta_0$ and $\{ \alpha_1, \beta_1, \gamma_1, \delta_1 \}$ up to an order of $O(\varepsilon^2)$ contribution.
Explicitly,
\begin{align}
    [\hat{\mathcal{H}}_{\delta_0}, \hat{\mathcal{H}}_{\alpha_1}] = - [\hat{\mathcal{H}}_{\alpha_1}, \hat{\mathcal{H}}_{\delta_0}] &=
    \begin{pmatrix}
    0 & 0 & 0 \\
    0 & 0 & - \frac{1}{2} a \\
    0 & 0 & 0 \\
    \end{pmatrix}
    , \\
    [\hat{\mathcal{H}}_{\delta_0}, \hat{\mathcal{H}}_{\beta_1}] = - [\hat{\mathcal{H}}_{\beta_1}, \hat{\mathcal{H}}_{\delta_0}] &=
    \begin{pmatrix}
    0 & 0 & 0 \\
    0 & 0 & \frac{1}{2} a \\
    0 & 0 & 0 \\
    \end{pmatrix}
    , \\
    [\hat{\mathcal{H}}_{\delta_0}, \hat{\mathcal{H}}_{\gamma_1}] = - [\hat{\mathcal{H}}_{\gamma_1}, \hat{\mathcal{H}}_{\delta_0}] &=
    \begin{pmatrix}
    0 & 0 & \frac{1}{2} a \\
    0 & 0 & 0 \\
    0 & 0 & 0 \\
    \end{pmatrix}
    , \\
    [\hat{\mathcal{H}}_{\delta_0}, \hat{\mathcal{H}}_{\delta_1}] = - [\hat{\mathcal{H}}_{\delta_1}, \hat{\mathcal{H}}_{\delta_0}] &=
    \begin{pmatrix}
    0 & 0 & \frac{1}{2} a \\
    0 & 0 & 0 \\
    0 & 0 & 0 \\
    \end{pmatrix}
    .
\end{align}
Moreover, from the structure of the velocity obtained in the last section [Eqs.~\eqref{eq:U_x}, \eqref{eq:U_y}, and \eqref{eq:Omega}], $\hat{\mathcal{F}}_{\mu \nu}$ in general has nonzero value only when the difference of the Fourier modes (i.e., the difference of $n$) of $\mu$ and $\nu$ is $0, \pm 1$.
By carefully evaluating all the quantities using Eqs.~\eqref{eq:U_x}, \eqref{eq:U_y}, and \eqref{eq:Omega}, we finally obtain the averaged velocity as
\begin{align}
    \langle U_x \rangle &= \frac{a}{8} \Bigg( 2 ( \langle \alpha_0 \dot{\alpha}_1 - \dot{\alpha}_0 \alpha_1 \rangle + \langle \alpha_0 \dot{\beta}_1 - \dot{\alpha}_0 \beta_1 \rangle ) \nonumber \\
    &\quad + \sum_{n=1}^{N-1} \bigg[ (2n-3) ( - \langle \alpha_n \dot{\alpha}_{n+1} - \dot{\alpha}_n \alpha_{n+1} \rangle - \langle \gamma_n \dot{\gamma}_{n+1} - \dot{\gamma}_n \gamma_{n+1} \rangle ) \nonumber \\
    &\qquad\quad + (2n-1) ( \langle \beta_n \dot{\alpha}_{n+1} - \dot{\beta}_n \alpha_{n+1} \rangle - \langle \delta_n \dot{\gamma}_{n+1} - \dot{\delta}_n \gamma_{n+1} \rangle ) \nonumber \\
    &\qquad\quad + (2n+1) ( \langle \alpha_n \dot{\beta}_{n+1} - \dot{\alpha}_n \beta_{n+1} \rangle + \langle \beta_n \dot{\beta}_{n+1} - \dot{\beta}_n \beta_{n+1} \rangle - \langle \gamma_n \dot{\delta}_{n+1} - \dot{\gamma}_n \delta_{n+1} \rangle + \langle \delta_n \dot{\delta}_{n+1} - \dot{\delta}_n \delta_{n+1} \rangle ) \bigg] \Bigg) + O(\varepsilon^3) ,
    \label{eq:U_x_final} \\
    \langle U_y \rangle &= \frac{a}{8} \Bigg( 2 ( \langle \alpha_0 \dot{\gamma}_1 - \dot{\alpha}_0 \gamma_1 \rangle - \langle \alpha_0 \dot{\delta}_1 - \dot{\alpha}_0 \delta_1 \rangle ) \nonumber \\
    &\quad + \sum_{n=1}^{N-1} \bigg[ (2n-3) ( - \langle \alpha_n \dot{\gamma}_{n+1} - \dot{\alpha}_n \gamma_{n+1} \rangle + \langle \gamma_n \dot{\alpha}_{n+1} - \dot{\gamma}_n \alpha_{n+1} \rangle ) \nonumber \\
    &\qquad\quad + (2n-1) ( \langle \delta_n \dot{\alpha}_{n+1} - \dot{\delta}_n \alpha_{n+1} \rangle + \langle \beta_n \dot{\gamma}_{n+1} - \dot{\beta}_n \gamma_{n+1} \rangle ) \nonumber \\
    &\qquad\quad + (2n+1) ( - \langle \alpha_n \dot{\delta}_{n+1} - \dot{\alpha}_n \delta_{n+1} \rangle - \langle \gamma_n \dot{\beta}_{n+1} - \dot{\gamma}_n \beta_{n+1} \rangle - \langle \beta_n \dot{\delta}_{n+1}  - \dot{\beta}_n \delta_{n+1} \rangle + \langle \delta_n \dot{\beta}_{n+1} - \dot{\delta}_n \beta_{n+1} \rangle ) \bigg] \Bigg) + O(\varepsilon^3) , \label{eq:U_y_final}\\
    \langle \Omega \rangle &= \frac{1}{2} \sum_{n=1}^N \bigg[ - (n-2) \langle \alpha_n \dot{\gamma}_n -  \dot{\alpha}_n \gamma_n \rangle - n ( \langle \alpha_n \dot{\delta}_n - \dot{\alpha}_n  \delta_n \rangle - \langle \beta_n \dot{\gamma}_n - \dot{\beta}_n \gamma_n \rangle + \langle \beta_n \dot{\delta}_n - \dot{\beta}_n \delta_n \rangle ) \bigg] + O(\varepsilon^3) .
    \label{eq:Omega_final}
\end{align}
The explicit forms of $\mathcal{U}_n^{\sigma,\sigma',i}$ and $\mathcal{W}_n^{\sigma,\sigma'}$ in Eqs.~\eqref{eq:U_i_main} and \eqref{eq:Omega_main} follow directly from Eqs.~\eqref{eq:U_x_final}-\eqref{eq:Omega_final}.

We can simplify this formula further to consider the case where only two modes out of all $\alpha_n, \beta_n, \gamma_n, \delta_n$ with $n>1$ are non-zero.
In the coefficient tables presented below, the columns are set as $\eta(t)=\cos (2\pi t/\tau)$, and the rows are $\xi(t)=\sin (2\pi t/\tau)$. Since
\begin{align}
\int_0 ^\tau [\eta(t)\dot{\xi}(t) - \dot{\eta}(t)\xi(t)] dt = 2\pi,
\end{align}
the contributions for these pairs of modes to the velocity and the rotational values will be $2\pi$ times these coefficients.

\begin{center}
\textbf{Table: $\langle U_x \rangle$}
\vspace{0.5em}

\begin{tabular}{l|cccc}
 & $\alpha_n$ & $\beta_n$ & $\gamma_n$ & $\delta_n$ \\ \hline
$\alpha_{n+1}$ & $-\frac{(2n-3)a}{8}$ & $+\frac{(2n-1)a}{8}$ & $0$ & $0$ \\
$\beta_{n+1}$  & $+\frac{(2n+1)a}{8}$ & $+\frac{(2n+1)a}{8}$ & $0$ & $0$ \\
$\gamma_{n+1}$ & $0$ & $0$ & $-\frac{(2n-3)a}{8}$ & $-\frac{(2n-1)a}{8}$ \\
$\delta_{n+1}$ & $0$ & $0$ & $-\frac{(2n+1)a}{8}$ & $+\frac{(2n+1)a}{8}$ \\
\end{tabular}

\vspace{1.2em}

\textbf{Table: $\langle U_y \rangle$}
\vspace{0.5em}

\begin{tabular}{l|cccc}
 & $\alpha_n$ & $\beta_n$ & $\gamma_n$ & $\delta_n$ \\ \hline
$\alpha_{n+1}$ & $0$ & $0$ & $+\frac{(2n-3)a}{8}$ & $+\frac{(2n-1)a}{8}$ \\
$\beta_{n+1}$  & $0$ & $0$ & $-\frac{(2n+1)a}{8}$ & $+\frac{(2n+1)a}{8}$ \\
$\gamma_{n+1}$ & $-\frac{(2n-3)a}{8}$ & $+\frac{(2n-1)a}{8}$ & $0$ & $0$ \\
$\delta_{n+1}$ & $-\frac{(2n+1)a}{8}$ & $-\frac{(2n+1)a}{8}$ & $0$ & $0$ \\
\end{tabular}

\vspace{1.2em}

\textbf{Table: $\langle \Omega \rangle$}
\vspace{0.5em}

\begin{tabular}{l|cccc}
 & $\alpha_n$ & $\beta_n$ & $\gamma_n$ & $\delta_n$ \\
\hline
$\alpha_{n}$ & $0$ & $0$ & \textbf{$+\frac{n-2}{2}$} & \textbf{$+\frac{n}{2}$} \\
$\beta_{n}$  & $0$ & $0$ & \textbf{$-\frac{n}{2}$} & \textbf{$+\frac{n}{2}$} \\
$\gamma_{n}$ & \textbf{$-\frac{n-2}{2}$} & \textbf{$+\frac{n}{2}$} & $0$ & $0$ \\
$\delta_{n}$ & \textbf{$-\frac{n}{2}$} & \textbf{$-\frac{n}{2}$} & $0$ & $0$ \\
\end{tabular}
\end{center}

\subsection{Swimming power}

To evaluate the swimming efficiency of the swimmer, we first derive the power (rate of work) needed for the given stroke.

We rewrite Eq.~\eqref{eq:Stokes} in polar coordinates as
\begin{equation}
    \left( \mathbf{e}_r \frac{\partial}{\partial r} + \mathbf{e}_\theta \frac{1}{r} \frac{\partial}{\partial \theta} \right) p = \mu \left( \frac{\partial^2}{\partial r^2} + \frac{1}{r} \frac{\partial}{\partial r} + \frac{1}{r^2} \frac{\partial^2}{\partial \theta^2} \right) (u \mathbf{e}_r + v \mathbf{e}_\theta) .
    \label{eq:Stokes_polar}
\end{equation}
By substituting Eqs.~\eqref{eq:u_A-D} and \eqref{eq:v_A-D} into Eq.~\eqref{eq:Stokes_polar}, and using the relations,
\begin{align}
    \frac{\partial}{\partial r} \mathbf{e}_r = \frac{\partial}{\partial r} \mathbf{e}_\theta &= 0 , \\
    \frac{\partial}{\partial \theta} \mathbf{e}_r &= \mathbf{e}_\theta , \\
    \frac{\partial}{\partial \theta} \mathbf{e}_\theta &= - \mathbf{e}_r ,
\end{align}
we find the pressure as
\begin{equation}
    p = 2 \mu \sum_{n=2}^\infty (n-1) \frac{a^{n-1}}{r^n} \left[ (A_n - B_n) \cos{n\theta} + (C_n + D_n) \sin{n\theta} \right]
    \label{eq:pressure}
\end{equation}
up to a constant of integration.
The surface stress tensor is given by
\begin{equation}
    \sigma_{i j} = p \delta_{i j} - 2 \mu e_{i j} ,
\end{equation}
where $\delta_{i j}$ is the Kronecker delta and $e_{i j}$ is the rate of strain tensor, whose components are explicitly given as
\begin{align}
    e_{r r} &= \frac{\partial u}{\partial r} , \\
    e_{\theta \theta} &= \frac{1}{r} \frac{\partial v}{\partial \theta} + \frac{u}{r} , \\
    e_{r \theta} &= \frac{1}{2} \left( \frac{\partial v}{\partial r} - \frac{v}{r} + \frac{1}{r} \frac{\partial u}{\partial \theta} \right) .
    \label{eq:e_theta_theta}
\end{align}
Then the surface stress is determined from Eqs.~\eqref{eq:u_A-D}, \eqref{eq:v_A-D}, and \eqref{eq:pressure}-\eqref{eq:e_theta_theta} as
\begin{align}
    \sigma_{r r} &= p - 2 \mu \frac{\partial u}{\partial r} \\
    &= 2 \mu \left\{ \frac{aA_0}{r^2} + (A_1 + B_1) \frac{a^2}{r^3} \cos{\theta} - (D_1 - C_1) \frac{a^2}{r^3} \sin{\theta} \right. \nonumber \\
    &\quad \qquad + \sum_{n=2}^\infty \frac{1}{2} A_n \cos{n\theta} \left[ - (n-2)(n+1) \frac{a^{n+1}}{r^{n+2}} + (n-1)(n+2) \frac{a^{n-1}}{r^n} \right] \nonumber \\
    &\quad \qquad + \sum_{n=2}^\infty \frac{1}{2} B_n \cos{n\theta} \left[ n(n+1) \frac{a^{n+1}}{r^{n+2}} - (n-1)(n+2) \frac{a^{n-1}}{r^n} \right] \nonumber \\
    &\quad \qquad + \sum_{n=2}^\infty \frac{1}{2} C_n \sin{n\theta} \left[ - (n-2)(n+1) \frac{a^{n+1}}{r^{n+2}} + (n-1)(n+2) \frac{a^{n-1}}{r^n} \right] \nonumber \\
    &\quad \qquad \left. + \sum_{n=2}^\infty \frac{1}{2} D_n \sin{n\theta} \left[ - n(n+1) \frac{a^{n+1}}{r^{n+2}} + (n-1)(n+2) \frac{a^{n-1}}{r^n} \right] \right\} , \\
    \sigma_{\theta \theta} &= p - 2 \mu \left( \frac{1}{r} \frac{\partial v}{\partial \theta} + \frac{u}{r} \right) \\
    &= - 2 \mu \left\{ \frac{aA_0}{r^2} + (A_1 + B_1) \frac{a^2}{r^3} \cos{\theta} - (D_1 - C_1) \frac{a^2}{r^3} \sin{\theta} \right. \nonumber \\
    &\quad \qquad + \sum_{n=2}^\infty \frac{1}{2} A_n \cos{n\theta} \left[ - (n-2)(n+1) \frac{a^{n+1}}{r^{n+2}} + (n-2)(n-1) \frac{a^{n-1}}{r^n} \right] \nonumber \\
    &\quad \qquad + \sum_{n=2}^\infty \frac{1}{2} B_n \cos{n\theta} \left[ n(n+1) \frac{a^{n+1}}{r^{n+2}} - (n-2)(n-1) \frac{a^{n-1}}{r^n} \right] \nonumber \\
    &\quad \qquad + \sum_{n=2}^\infty \frac{1}{2} C_n \sin{n\theta} \left[ - (n-2)(n+1) \frac{a^{n+1}}{r^{n+2}} + (n-2)(n-1) \frac{a^{n-1}}{r^n} \right] \nonumber \\
    &\quad \qquad \left. + \sum_{n=2}^\infty \frac{1}{2} D_n \sin{n\theta} \left[ - n(n+1) \frac{a^{n+1}}{r^{n+2}} + (n-2)(n-1) \frac{a^{n-1}}{r^n} \right] \right\} , \\
    \sigma_{r \theta} &= - \mu \left( \frac{\partial v}{\partial r} - \frac{v}{r} + \frac{1}{r} \frac{\partial u}{\partial \theta} \right) \\
    &= 2 \mu \left\{ (A_1 + B_1) \frac{a^2}{r^3} \sin{\theta} + (D_1 - C_1) \frac{a^2}{r^3} \cos{\theta} \right. \nonumber \\
    &\quad \qquad + \sum_{n=2}^\infty \frac{1}{2} A_n \sin{n\theta} \left[ - (n-2)(n+1) \frac{a^{n+1}}{r^{n+2}} + n(n-1) \frac{a^{n-1}}{r^n} \right] \nonumber \\
    &\quad \qquad + \sum_{n=2}^\infty \frac{1}{2} B_n \sin{n\theta} \left[ n(n+1) \frac{a^{n+1}}{r^{n+2}} - n(n-1) \frac{a^{n-1}}{r^n} \right] \nonumber \\
    &\quad \qquad + \sum_{n=2}^\infty \frac{1}{2} C_n \cos{n\theta} \left[ (n-2)(n+1) \frac{a^{n+1}}{r^{n+2}} - n(n-1) \frac{a^{n-1}}{r^n} \right] \nonumber \\
    &\quad \qquad \left. + \sum_{n=2}^\infty \frac{1}{2} D_n \cos{n\theta} \left[ n(n+1) \frac{a^{n+1}}{r^{n+2}} - n(n-1) \frac{a^{n-1}}{r^n} \right] \right\} .
\end{align}
Therefore, we have the rate of work per unit area, $P$, as
\begin{align}
    P &= \frac{1}{2\pi} \int_{0}^{2\pi} (u \sigma_{r r} + v \sigma_{r \theta})|_{r=a} d\theta \\
    &= \frac{\mu}{a} \left\{ 2 A_0^2 + \sum_{n=1}^\infty \left[ n (A_n^2 + B_n^2 + C_n^2 + D_n^2) + 2 A_n B_n - 2 C_n D_n \right] \right\} .
\end{align}
From Eqs.~\eqref{eq:A_n_first}-\eqref{eq:D_n_first}, at the leading order, $O(\varepsilon^2)$, we have
\begin{equation}
    P = \mu a \left\{ 2 \dot{\alpha}_0^2 + \sum_{n=1}^\infty \left[ n (\dot{\alpha}_n^2 + \dot{\beta}_n^2 + \dot{\gamma}_n^2 + \dot{\delta}_n^2) + 2 \dot{\alpha}_n \dot{\beta}_n - 2 \dot{\gamma}_n \dot{\delta}_n \right] \right\} + O(\varepsilon^3) .
\end{equation}
Upon time averaging, we obtain
\begin{equation}
    \langle P \rangle = \mu a \left\{ 2 \langle \dot{\alpha}_0^2 \rangle + \sum_{n=1}^\infty \left[ n ( \langle \dot{\alpha}_n^2 \rangle + \langle \dot{\beta}_n^2 \rangle + \langle \dot{\gamma}_n^2 \rangle + \langle \dot{\delta}_n^2 \rangle ) + 2 \langle \dot{\alpha}_n \dot{\beta}_n \rangle - 2\langle  \dot{\gamma}_n \dot{\delta}_n \rangle \right] \right\} + O(\varepsilon^3) .
\end{equation}

\subsection{Efficiency bound by symmetrized strokes}
Let us consider the specific case where there is no rotation. Without loss of generality, we can set the $x-$axis as the direction of swimming, in which case $\langle U_y \rangle =0$. Let us also focus on the case where there are only modes that are $n>1$.
For any swimming strategy under this constraint, we can symmetrize it with respect to the $x-$axis by setting 
\begin{align}
    {\alpha}_n^\pm &= \alpha_n \pm \gamma_n \\
    {\beta}_n^\pm &= \beta_n \mp \delta_n,
\end{align}
and $\gamma_n^\pm = \delta^\pm_n = 0$.
Let us denote the velocity in the $x-$directions calculated using these factors as ${U}^\pm_x$ and the rate of works as $\langle {P^\pm}\rangle$.
Since this swimming strategy is symmetric with respect to the $x-$axis, we have ${U}^\pm_y=0$. Now, from the formula we can easily show that the following holds:
\begin{align}
    \langle {U}^+_x\rangle + \langle {U}^-_x\rangle &=  2 \langle {U}_x\rangle  \\
    \langle {P}^+\rangle + \langle {P}^-\rangle &=  2 \langle {P}\rangle.
\end{align}

Let us further define the efficiencies of swimming by $\eta=|\langle {U}_x\rangle|/\langle {P}\rangle$ and $\eta^\pm=|\langle {U}^\pm_x\rangle|/\langle {P^\pm}\rangle$. 
Applying the triangle inequality, $|\langle U_x^+ \rangle + \langle U_x^- \rangle| \le |\langle U_x^+ \rangle| + |\langle U_x^- \rangle|$,
\begin{align}
\eta = \frac{|\langle U_x \rangle|}{\langle P \rangle} = \frac{|\langle U_x^+ \rangle + \langle U_x^- \rangle|}{\langle P^+ \rangle + \langle P^- \rangle} \le \frac{|\langle U_x^+ \rangle| + |\langle U_x^- \rangle|}{\langle P^+ \rangle + \langle P^- \rangle} = \frac{\eta^+ \langle P^+ \rangle + \eta^- \langle P^- \rangle}{\langle P^+ \rangle + \langle P^- \rangle}.
\end{align}
The final expression is a weighted average of the two symmetrized efficiencies, with the non-negative powers $\langle P^+ \rangle$ and $\langle P^- \rangle$ as the weights. Any weighted average is necessarily bounded by the maximum of its components. Thus, we arrive at the result
\begin{align}
\eta \le \max(\eta^+, \eta^-).
\end{align}
This proves that the efficiency of a non-symmetric stroke cannot exceed that of the more efficient of its two symmetric counterparts. Consequently, the search for an optimal swimming strategy, in terms of efficiency, can be restricted to the subspace of purely symmetric deformations. Note that $\eta \ge \min(\eta^+, \eta^-)$ does not generally hold.

\subsection{Non-symmetric swimmers from matched symmetric--anti-symmetric superposition}
\label{sec:nonsym_construction}

Given any symmetric swimmer with deformation amplitudes $(\alpha_n, \beta_n)$ and $\gamma_n = \delta_n = 0$, one can construct a one-parameter family of non-symmetric swimmers by adding the anti-symmetric dual stroke with the same shape and phase but scaled by a real parameter $\lambda$:
\begin{equation}
\gamma_m = \lambda\,\alpha_m, \qquad \delta_m = -\lambda\,\beta_m.
\label{eq:nonsym_construction}
\end{equation}
This superposition preserves the swimming direction: $\langle \Omega \rangle = 0$ and $\langle U_y \rangle = 0$, as can be verified by substituting into Eqs.~\eqref{eq:Omega_final} and \eqref{eq:U_y_final}. 

Since the two symmetrized components are $\bm{v}^\pm = (1 \pm \lambda)\bm{v}$, where $\bm{v}$ is the original symmetric stroke, both yield the same efficiency $\eta^+ = \eta^- = \eta$. It follows that the efficiency bound of the previous subsection is tight: if the original symmetric swimmer achieves $\eta^\ast$, the non-symmetric family achieves $\eta^\ast$ as well.

Note that this is a specific family of non-symmetric swimmers, which is described by the sum of a symmetric and an anti-symmetric stroke with matched shape and phase. In general, when the symmetric and anti-symmetric components have different shapes or phases, the mismatch generates net rotation ($\langle \Omega \rangle \neq 0$).

\subsection{Optimal strokes}

Here we analyze optimal strokes with a focus on symmetrized  deformations. The swimming speed and power consumption are written in the forms, 
\begin{align}
    \langle U_x\rangle=\frac{1}{\tau}\int_0^\tau\sum_{m,n,\sigma}\mathcal{F}_{mn}^\sigma\sigma_m\dot{\sigma}_ndt,~\textrm{and}~\langle P\rangle=\frac{1}{\tau}\int_0^\tau\sum_{m,n,\sigma}\mathcal{P}_{mn}^\sigma\dot{\sigma}_m\dot{\sigma}_n dt,
\end{align}
where $\sigma\in\{\alpha, \beta \}$. Following \cite{shapere1989efficiencies}, we consider a variation $\delta \eta=0$, which leads to the variation equation, $\delta\langle U_x \rangle=\eta \delta\langle P\rangle$. With antisymmetric and symmetric properties, $\mathcal{F}^\sigma_{mn}=-\mathcal{F}^\sigma_{nm}$ and $\mathcal{P}^\sigma_{mn}=\mathcal{P}^\sigma_{nm}$, we may obtain
\begin{align}
    \sum_{m,n,\sigma}\mathcal{F}^\sigma_{mn}\dot{\sigma}_n=\eta \sum_{m,n,\sigma}\mathcal{P}^\sigma_{mn}\ddot{\sigma}_n.
\end{align}
We then expand the deformation in a Fourier series in time as $\bm{\sigma}=\bm{\sigma}_0e^{i\omega t}$ with  $\bm{\sigma}_0$ being its mean value and we will obtain the optimal efficiency $\eta$ as a function of $\omega$ from the eigenvalue problem,
\begin{align}
    {\bf A}\bm{\sigma}_0=\eta\bm{\sigma}_0,
    \label{eq:eigprob}
\end{align}
where the matrix {\bf A} is given by ${\bf A}=-(i/\omega){\bf P}^{-1}{\bf F}$ with ${\bf P}$ and ${\bf F}$ being the matrix representation of $\mathcal{P}^\sigma_{mn}$ and $\mathcal{F}^\sigma_{mn}$, respectively.

As an illustrative example, we consider a stroke with two dominating modes $n$ and $n+1$ ($n\geq 2$) so that remaining modes vanish as $\sigma_m=0$ for $m\neq n, n+1$. Let $\bm{\sigma}_0=(\alpha_n, \beta_n, \alpha_{n+1}, \beta_{n+1})$ and solve the four-dimensional eigenvalue problem \eqref{eq:eigprob}. The maximum eigenvalue is then obtained as

\begin{equation}
\eta^\ast _n = \frac{1}{16\pi} \sqrt{ \frac{2 \left( \sqrt{\Gamma(n)} + \Phi(n) \right)}{n(n+1)(n+2)} },
\end{equation}
where the polynomials $\Gamma(n)$ and $\Phi(n)$ are defined as
\begin{align}
\Gamma(n) &= 8n^4 + 12n^3 + 13n^2 + 12n + 4 \\
\Phi(n) &= 4n^3 + 6n^2 - n - 2.
\end{align}
We also obtain the second eigenvalue, which is, however, smaller than $\eta^\ast_n$.

The corresponding eigenvectors to the eigenvalue $\eta^\ast _n$ are explicitly given by
\begin{equation}
\left(\begin{array}{c} 
\pm 4\pi i {\eta^\ast}\left[ (3n+2)(6n^2+3n-2)-(2n^2+n+2)\sqrt{\Gamma(n)}
\right]\\ 
\mp 4\pi i {\eta^\ast}\left[ 8n^4+14n^3+5n^2+4n+4-(2n^2+5n-2)\sqrt{\Gamma(n)}
\right]
\\
(2n^2-n-1)\left[ 2(n+1)^2-\sqrt{\Gamma(n)}\right]
\\ (2n^2-3n-4)(2n^2-n-1) \end{array}\right).
\end{equation}
For $n \to \infty$, the efficiency converges to 
\begin{equation}
\eta^\ast_\infty = \frac{\sqrt{2}}{8\pi} \approx 0.0563.
\end{equation}

When $n=2$, the maximum efficiency is explicitly given by 
\begin{equation}
  \eta^\ast_2 = \frac{\sqrt{3}}{48\pi} \sqrt{13 + \sqrt{19}} \approx 0.0479,
\label{eq:eta_star}
\end{equation}
and the corresponding eigenvector is
\begin{equation}
    \begin{pmatrix} 
\frac{2\sqrt{3}}{15} i \left(3\sqrt{19}-14\right)\sqrt{13+\sqrt{19}} \\ 
-\frac{2\sqrt{3}}{15}i\left(4\sqrt{19}-17\right)\sqrt{13+\sqrt{19}} \\ 
2\sqrt{19}-9 \\ 
1 
\end{pmatrix}
\approx
\begin{pmatrix} -0.8884 i \\ -0.4191 i \\ -0.2822 \\ 1 \end{pmatrix}.
\end{equation}
The four components correspond to $(\alpha_2,\beta_2,\alpha_3,\beta_3)$ amplitudes; the interpretation of $i$ is that the first two modes oscillate as $\sin(\omega t)$ and the latter two as $\cos(\omega t)$, giving a 90$^\circ$ phase offset between mode-$2$ and mode-$3$ deformations.

When $n=3$, the maximum efficiency is
\begin{equation}
  \eta^\ast_3 = \frac{1}{16\pi} \sqrt{\frac{\sqrt{1129} + 157}{30}} \approx 0.0501,
\label{eq:eta_star_n3}
\end{equation}
and the corresponding eigenvector is
\begin{equation}
    \begin{pmatrix}
-\dfrac{23\sqrt{1129}-671}{280} i \sqrt{\dfrac{\sqrt{1129}+157}{30}} \\[8pt]
-\dfrac{1087-31\sqrt{1129}}{280} i \sqrt{\dfrac{\sqrt{1129}+157}{30}} \\[8pt]
\dfrac{32-\sqrt{1129}}{5} \\[6pt]
1
\end{pmatrix}
\approx
\begin{pmatrix} -0.9165 i \\ -0.4085 i \\ -0.3201 \\ 1 \end{pmatrix}.
\label{eq:evec_n3}
\end{equation}
The four components correspond to $(\alpha_3,\beta_3,\alpha_4,\beta_4)$ amplitudes; the first two modes oscillate as $\sin(\omega t)$ and the latter two as $\cos(\omega t)$.

\section{Numerical method for calculating swimming velocity at low Reynolds number}

We computed the swimming motion of deforming bodies at low Reynolds number using a boundary-integral formulation of the incompressible Stokes equations in both two and three dimensions.
In this regime, inertial effects are negligible and the fluid dynamics are governed by
\begin{align}
-\nabla p + \mu \nabla^2 \mathbf{u} &= \mathbf{0}, \\
\nabla \cdot \mathbf{u} &= 0,
\end{align}
where $\mathbf{u}$ is the fluid velocity, $p$ the pressure, and $\mu$ the dynamic viscosity.
The swimmer is assumed to be force- and torque-free at all times.

\subsection{Boundary integral formulation (2D and 3D)}

In an unbounded domain, Stokes flow can be represented by a single-layer potential on the swimmer boundary,
\begin{align}
\mathbf{u}(\mathbf{x})
=
\int_{\partial\Omega}
\mathbf{G}(\mathbf{x}-\mathbf{y})\mathbf{f}(\mathbf{y}) \mathrm{d}S(\mathbf{y}),
\end{align}
where $\partial\Omega$ denotes the swimmer boundary, $\mathbf{f}(\mathbf{y})$ is the (unknown) surface force density exerted by the swimmer on the fluid, and $\mathbf{G}$ is the Stokeslet Green's function \cite{pozrikidis1992boundary}.
In two dimensions, $\mathrm{d}S$ reduces to the arc-length element $\mathrm{d}s$ and $\mathbf{G}$ is the two-dimensional Stokeslet.
In three dimensions, $\mathrm{d}S$ is the surface-area element and $\mathbf{G}$ is the three-dimensional Stokeslet.

To avoid numerical singularities associated with point forces, we employed a regularized Stokeslet formulation \cite{cortez2001method, cortez2005method3d}, using a regularization length $\delta$ chosen proportional to the local discretization scale.
In two dimensions we used
\begin{align}
\mathbf{G}_\delta^{\mathrm{2D}}(\mathbf{r})
=
-\frac{1}{4\pi\mu}
\left[
\frac{1}{2}\ln(r^2+\delta^2)\mathbf{I}
-
\frac{\mathbf{r}\mathbf{r}^{\mathsf T}}{r^2+\delta^2}
\right],
\end{align}
and in three dimensions
\begin{align}
\mathbf{G}_\delta^{\mathrm{3D}}(\mathbf{r})
=
\frac{1}{8\pi\mu}
\frac{(r^2+2\delta^2)\mathbf{I}+\mathbf{r}\mathbf{r}^{\mathsf T}}{(r^2+\delta^2)^{3/2}}.
\end{align}
This smoothing removes the kernel singularity while preserving the correct far-field Stokeslet behavior.

\subsection{Boundary discretization}

In 2D, the swimmer boundary was discretized into $M$ ordered material points
$\{\mathbf{r}_i(t)\}_{i=1}^M$ forming a closed polygonal approximation of $\partial\Omega$.
Each point carried a quadrature weight $w_i$ estimated from the local arc length,
\begin{align}
w_i \approx \frac{1}{2}\left(
\lvert \mathbf{r}_{i+1}-\mathbf{r}_i \rvert
+
\lvert \mathbf{r}_i-\mathbf{r}_{i-1} \rvert
\right),
\end{align}
with periodic indexing.
The regularization length was set as
\begin{align}
\delta = \varepsilon_{\mathrm{reg}} h, 
\qquad  h \equiv \frac{1}{M}\sum_{i=1}^{M} \lvert \mathbf{r}_{i+1}-\mathbf{r}_i \rvert,
\end{align}
where $h$ is the mean segment length.

In 3D, we discretized the swimmer surface using $N$ quasi-uniform points $\{\mathbf{x}_i(t)\}_{i=1}^N$ generated by a Fibonacci sphere rule and mapped to the instantaneous surface. Each point carried a quadrature weight proportional to the local area element on the spheroid,
\begin{align}
w_i = \frac{4\pi}{N}a\sqrt{b^2(1-\hat{n}_{z,i}^2) + a^2 \hat{n}_{z,i}^2},  
\end{align}                                                                                         where $\hat{n}_{z,i}$ is the $z$-component of the $i$-th reference point on the unit sphere and $a$, $b$ are the equatorial and polar semi-axes.
The regularization length was set as
\begin{align}
\delta = \varepsilon_{\mathrm{reg}} h,
\qquad
h \equiv \sqrt{\frac{S}{N}},
\end{align}
where $h$ is a representative point spacing and $\varepsilon_{\mathrm{reg}}$ is a dimensionless regularization parameter.

\subsection{Boundary kinematics and surface actuation}

At each boundary point, the no-slip condition requires the fluid velocity to match the surface velocity.
The instantaneous boundary velocity was decomposed as
\begin{align}
\mathbf{u}(\mathbf{x}_i)
=
\mathbf{U}
+
\boldsymbol{\Omega}\times(\mathbf{x}_i-\mathbf{x}_c)
+
\mathbf{v}_{\mathrm{bc}}(\mathbf{x}_i),
\end{align}
where $\mathbf{U}$ and $\boldsymbol{\Omega}$ denote the rigid-body translational and angular velocities, respectively, and $\mathbf{x}_c$ is the instantaneous center of mass (computed from the discretized surface).
The prescribed boundary velocity $\mathbf{v}_{\mathrm{bc}}$ was obtained directly from the time evolution of the material points.
In 2D this was implemented by finite differences,
\begin{align}
\mathbf{v}_{\mathrm{bc}}(\mathbf{r}_i)
=
\frac{\mathbf{r}_i(t+\Delta t)-\mathbf{r}_i(t)}{\Delta t},
\end{align}
which naturally incorporates both normal shape deformations and tangential motion of surface material points.
In 3D, $\mathbf{v}_{\mathrm{bc}}$ was computed analytically from the prescribed deformation protocol when available; for general user-defined time dependence of the deformation coefficients we used a centered finite difference in time.

\subsection{Force- and torque-free constraints}

Because the swimmer is self-propelled, the total hydrodynamic force and torque must vanish at every instant,
\begin{align}
\int_{\partial\Omega} \mathbf{f}(\mathbf{y}) \mathrm{d}S(\mathbf{y}) &= \mathbf{0}, \\
\int_{\partial\Omega} (\mathbf{y}-\mathbf{x}_c)\times \mathbf{f}(\mathbf{y}) \mathrm{d}S(\mathbf{y}) &= \mathbf{0}.
\end{align}
These conditions determine the unknown rigid-body velocities $\mathbf{U}$ and $\boldsymbol{\Omega}$.

\subsection{Discrete linear system}

After discretization, the boundary integral equation and the force- and torque-free constraints yield a linear system at each time step. In 3D, the discrete form is
\begin{align}
\sum_{j=1}^N
\mathbf{G}_\delta^{\mathrm{3D}}(\mathbf{x}_i-\mathbf{x}_j)
\mathbf{f}_j w_j
=
\mathbf{U}
+
\boldsymbol{\Omega}\times(\mathbf{x}_i-\mathbf{x}_c)
+
\mathbf{v}_{\mathrm{bc}}(\mathbf{x}_i),
\qquad i=1,\ldots,N,
\end{align}
supplemented by the discrete force- and torque-free constraints
\begin{align}
\sum_{j=1}^N \mathbf{f}_j w_j &= \mathbf{0}, \\
\sum_{j=1}^N \bigl[(\mathbf{x}_j-\mathbf{x}_c)\times \mathbf{f}_j \bigr] w_j &= \mathbf{0}.
\end{align}
The unknowns are the force densities $\{\mathbf{f}_j\}_{j=1}^N$, the translational velocity $\mathbf{U}$, and the angular velocity $\boldsymbol{\Omega}$.
The system can be solved using direct linear solvers.
(An analogous system was solved in 2D, with $\mathbf{U}\in\mathbb{R}^2$ and scalar angular velocity.)

Once $\mathbf{U}(t)$ and $\boldsymbol{\Omega}(t)$ were obtained, the swimmer position $\mathbf{X}(t)$ was updated by explicit time integration,
\begin{align}
\dot{\mathbf{X}}(t) &= \mathbf{U}(t).
\end{align}
For the 3D simulations, the visualized trajectory was reconstructed by translating the instantaneous body-frame point cloud by $\mathbf{X}(t)$; rotational effects can be incorporated by integrating $\dot{\mathbf{R}}=\boldsymbol{\Omega}\times\mathbf{R}$.

\subsection{Stroke protocols and sampling procedures}

We describe the specific stroke protocols used in the numerical simulations of Fig.~\ref{fig3}.

\vspace{2mm}

\noindent\textbf{Theory validation (Fig.~\ref{fig3}a):} Five stroke classes were tested, each consisting of a two-mode deformation protocol with sinusoidal time dependence at angular frequency $\omega = 2\pi/\tau$:
\begin{enumerate}
\item \emph{Symmetric $(n{=}2,3)$}: $\alpha_2(t) = \varepsilon \cos\omega t$, $\beta_3(t) = \varepsilon \sin\omega t$.
\item \emph{Symmetric $(n{=}3,4)$}: $\alpha_3(t) = \varepsilon \cos\omega t$, $\beta_4(t) = \varepsilon \sin\omega t$.
\item \emph{Anti-symmetric $(n{=}2,3)$}: $\gamma_2(t) = \varepsilon \cos\omega t$, $\delta_3(t) = \varepsilon \sin\omega t$.
\item \emph{Combined}: $\alpha_2(t) = \varepsilon \cos\omega t$, $\beta_3(t) = \varepsilon \sin\omega t$, $\gamma_2(t) = \varepsilon \sin\omega t$, $\delta_3(t) = \varepsilon \cos\omega t$. This stroke simultaneously activates symmetric and anti-symmetric modes, and also induce rotation.
\item \emph{Rotational}: $\alpha_3(t) = \varepsilon \cos\omega t$, $\gamma_3(t) = \varepsilon \sin\omega t$. 
\end{enumerate}
All other coefficients are zero. The first two strokes are symmetric, the third is anti-symmetric, the fourth is non-symmetric (mixing both symmetry classes), and the fifth generates pure rotation via same-order mode coupling. For each stroke class, the ratio of numerically computed to analytically predicted swimming speed was evaluated over $16$ logarithmically spaced values of $\varepsilon$ from $0.01$ to $0.36$, using $M=1152$ boundary vertices and $T=1024$ time steps per period.

\vspace{2mm}

\noindent\textbf{Duality test with random strokes (Fig.~\ref{fig3}b):}
To test hydrodynamic duality beyond the five specific strokes above, we generated random multi-mode stroke pairs as follows. For each sample, three wavenumbers were drawn uniformly without replacement from $\{2,3,4,5\}$. For each active wavenumber $n$, normal and tangential amplitudes $A_n$ and $B_n$ were drawn from $\mathrm{Uniform}(-1,1)/(n-1)$ (scaled by $1/(n-1)$ to moderate the amplitude of higher modes), and phases $\phi^A_n$, $\phi^B_n$ were drawn from $\mathrm{Uniform}(0,2\pi)$. The symmetric swimmer was constructed by assigning $\alpha_n(t) = \varepsilon A_n \cos(\omega t + \phi^A_n)$ and $\beta_n(t) = \varepsilon B_n \cos(\omega t + \phi^B_n)$; its anti-symmetric dual was obtained via the duality transformation $\alpha_n \to -\gamma_n$, $\beta_n \to \delta_n$.

In addition, the reference body was deformed from a circle to an ellipse by adding a static component $\alpha_2^{(0)} = \varepsilon_d/\varepsilon$ to the $n=2$ radial mode, where the ellipticity parameter $\varepsilon_d$ was drawn from $\mathrm{Uniform}(-0.2, 0.2)$. Samples were discarded if the boundary parameterisation became non-monotonic or developed cusps at any time during the swimming cycle, or if the computed speed was below $0.001\,a/\tau$, or if the lateral velocity exceeded $30\%$ of the forward velocity. In total, $256$ valid samples were retained from $328$ attempts. Each simulation used $M=576$ boundary vertices, $T=256$ time steps, and deformation amplitude $\varepsilon=0.15$.

\vspace{2mm}

\noindent\textbf{Three-dimensional swimmers (Fig.~\ref{fig3}d).}
Two stroke pairs were tested on the ellipsoid slice swimmer, each in symmetric and anti-symmetric variants:
\begin{itemize}
\item Pair~1 ($n{=}2,3$): symmetric $\alpha_2(t) = \varepsilon \cos\omega t$, $\beta_3(t) = \varepsilon \sin\omega t$; anti-symmetric $\gamma_2(t) = \varepsilon \cos\omega t$, $\delta_3(t) = \varepsilon \sin\omega t$.
\item Pair~2 ($n{=}3,4$): symmetric $\alpha_3(t) = \varepsilon \cos\omega t$, $\beta_4(t) = \varepsilon \sin\omega t$; anti-symmetric $\gamma_3(t) = \varepsilon \cos\omega t$, $\delta_4(t) = \varepsilon \sin\omega t$.
\end{itemize}
Each pair was simulated across seven body aspect ratios $L_z/L_x \in \{0.67, 0.80, 0.90, 1.00, 1.10, 1.25, 1.50\}$ at deformation amplitude $\varepsilon = 0.15$, using $N=480$ surface points and $T=64$ time steps per period.

\subsection{Choice of the regularization length}
\label{subsec:reg_selection}

The regularized Stokeslet formulation introduces a smoothing length $\delta$ that must be chosen relative to the spatial discretization.
A common practical guideline is to set $\delta$ proportional to a characteristic boundary spacing $h$, as described above.
While this scaling is widely used~\cite{cortez2005method3d, ainley2008images, smith2009boundary}, the optimal value of $\varepsilon_{\mathrm{reg}}$ is not universal and depends on the discretization, quadrature rule, and the specific regularized kernel.
We therefore determined $\varepsilon_{\mathrm{reg}}$ empirically using a mesh-consistency criterion.

For a given actuation protocol and fixed temporal resolution, we performed one-period simulations at two spatial resolutions (a coarse discretization $M_c$ and a finer discretization $M_f$) while scanning $\varepsilon_{\mathrm{reg}}$ over a prescribed range.
For each $\varepsilon_{\mathrm{reg}}$, we computed the period-averaged swimming velocity $\langle U_x\rangle$ at the two resolutions and selected $\varepsilon_{\mathrm{reg}}$ by minimizing the mismatch
\begin{align}
\left|
\frac{\langle U_x\rangle_{f}}{\langle U_x\rangle_{c}} - 1
\right|,
\end{align}
excluding parameter values for which the coarse-resolution velocity was negligibly small.
The selected $\varepsilon_{\mathrm{reg}}$ was then fixed and used in higher-resolution simulations, where we verified that the resulting swimming velocities and net displacements were insensitive to further increases in $N$.

\subsection{Validation and numerical parameters}

We validated the numerical scheme using small-amplitude actuation protocols for which analytical predictions are available.
In 2D, we compared the computed mean translational and rotational velocities for two-mode actuation protocols against analytical small-amplitude coefficients and verified the expected quadratic scaling with the deformation amplitude.

In 3D, we validated the implementation by computing swimming speeds for two independent stroke
pairs---$(n=2,3)$ and $(n=3,4)$ modes---each in symmetric and anti-symmetric variants, across
the body aspect ratios $L_z/L_x=b/a\in[0.67,1.50]$ and with deformation amplitude
$\varepsilon = 0.15$ (14 pairs of simulations total).
In all cases, symmetric and anti-symmetric strokes agreed to within $0.4\%$ across all tested amplitudes, providing numerical evidence for the duality $U_\mathrm{sym}\approx U_\mathrm{anti}$ and providing an internal consistency check of the 3D implementation across a broad range of parameters.

Numerically, the dimensionless regularization parameter $\varepsilon_{\mathrm{reg}}$ was selected by a coarse--fine consistency criterion: we scanned $\varepsilon_{\mathrm{reg}}$ and chose values for which the period-averaged velocity obtained at $M=36$ and $M=72$ surface points agreed to within a prescribed tolerance, and then verified stability of the result at $M=1728$.

Having validated the 3D implementation, we computed swimming velocities for the ellipsoid slice swimmer.
In this model, the swimmer surface was represented by quasi-uniform points on the unit sphere mapped to an ellipsoid-like body with semi-axes $L_x=L_y=a$ in the $x$--$y$ directions and a dorsal--ventral semi-axis $L_z=b$ along $z$.
Deformations were prescribed on each $z$-slice as a planar shape change in the $x$--$y$ plane: the local slice radius $R_0=a\sqrt{1-n_z^2}$ was modulated by a radial component $s_R(\theta,t)$ and a tangential reparameterization $s_T(\theta,t)$ expressed as truncated Fourier series up to mode $N_{\mathrm{def}}$,
\begin{align}
s_R(\theta,t) &= \sum_{n=0}^{N_{\mathrm{def}}} \bigl[\alpha_n(t)\cos(n\theta)+\gamma_n(t)\sin(n\theta)\bigr],\\
s_T(\theta,t) &= \sum_{n=0}^{N_{\mathrm{def}}} \bigl[\beta_n(t)\sin(n\theta)+\delta_n(t)\cos(n\theta)\bigr].
\end{align}
The instantaneous surface points and boundary velocities were then assembled into a three-dimensional point cloud and used as boundary data in the force- and torque-free boundary-integral solver, yielding $\mathbf{U}(t)$ and $\boldsymbol{\Omega}(t)$ and the net displacement per actuation period.
Production simulations used $N=480$ surface points and $T=64$ time steps per actuation period, with $\varepsilon_{\mathrm{reg}}$ chosen by the coarse--fine consistency procedure described above using coarse ($N=180$) and fine ($N=360$) resolutions.

\subsection{Comparison between theory and simulation}

The discrepancy between theoretical predictions and numerical results depends on the number of surface vertices $M$, the number of time steps per period $T$, and the regularization parameter $\varepsilon_{\mathrm{reg}}$. Empirically, we found that the relative error scales as
\begin{align}
\frac{ U_{\mathrm{num}} - U_{\mathrm{th}} }{ U_{\mathrm{th}} }
\approx
c_1(\varepsilon_{\mathrm{reg}}) M^{-\mu_1}
+
c_2 M^{-\mu_2}
+
c_3(\varepsilon_{\mathrm{reg}}) T^{-\nu}, \label{eq:scaling}
\end{align}
where all three coefficients also depend on the swimming mode. We observed $\mu_1 \sim 0.9$ and $\mu_2 \sim 0.5$, implying that the first term in Eq.~(\ref{eq:scaling}) decays more rapidly with spatial resolution than the second. We also observed $\nu \sim 1.0$, which was independent of the time-integration scheme (e.g., forward Euler vs. fourth-order Runge--Kutta). 

The second and third terms dominate at large $M$. However, we find that the coefficient $c_1$ depends on the regularization parameter, and if $c_1(\varepsilon_{\mathrm{reg}}) > c_2$ then for small $M$ the first term can be more dominant. We noticed that $c_1(\varepsilon_{\mathrm{reg}})$ can be minimized by optimizing $\varepsilon_{\mathrm{reg}}$; employing the mesh-consistency criterion described in Sec.~\ref{subsec:reg_selection}, we found that $c_1$ can be effectively eliminated ($c_1 \approx 0$) for certain shape configurations, although this cancellation is not universal across all swimming modes, which seems to be why the pure rotation case tends to deviate largest from theory  (Fig.~\ref{fig3}a, Extended Data Fig.~\ref{figS2}). In regimes where the first term is successfully suppressed, the error becomes dominated by the second term even at relatively moderate spatial resolutions (Extended Data Fig.~\ref{figS3}).

\clearpage
\newpage

\renewcommand{\figurename}{Extended Data Fig.}
\setcounter{figure}{0} 

\begin{figure*}[h!]
\centering
\includegraphics[width=3.25in]{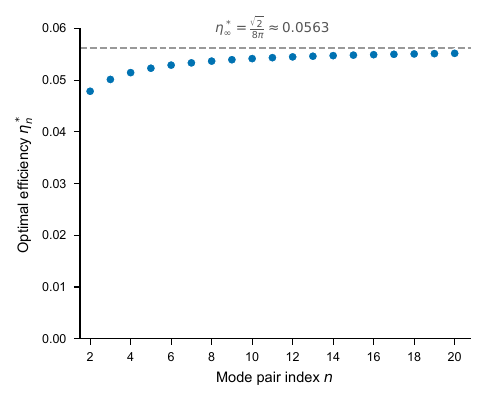}
\caption{\textbf{Optimal swimming efficiency as a function of mode number.}
Maximum swimming efficiency $\eta^*_n$ for a two-mode stroke using Fourier modes $n$ and $n{+}1$, computed from the eigenvalue problem (Eq.~\eqref{eq:eigprob}). The efficiency increases monotonically with mode number and converges to the limiting value $\eta^*_\infty = \sqrt{2}/(8\pi) \approx 0.0563$ (dashed line), first obtained by Shapere and Wilczek~\cite{shapere1989efficiencies}. Closed-form expressions for $\eta^*_2$ and $\eta^*_3$ are given in Eqs.~\eqref{eq:eta_star} and \eqref{eq:eta_star_n3}. By hydrodynamic duality, the same efficiency values apply to the corresponding anti-symmetric strokes.
}
\label{figS1}
\end{figure*}

\begin{figure*}[h!]
\centering
\includegraphics[width=1.0\linewidth]{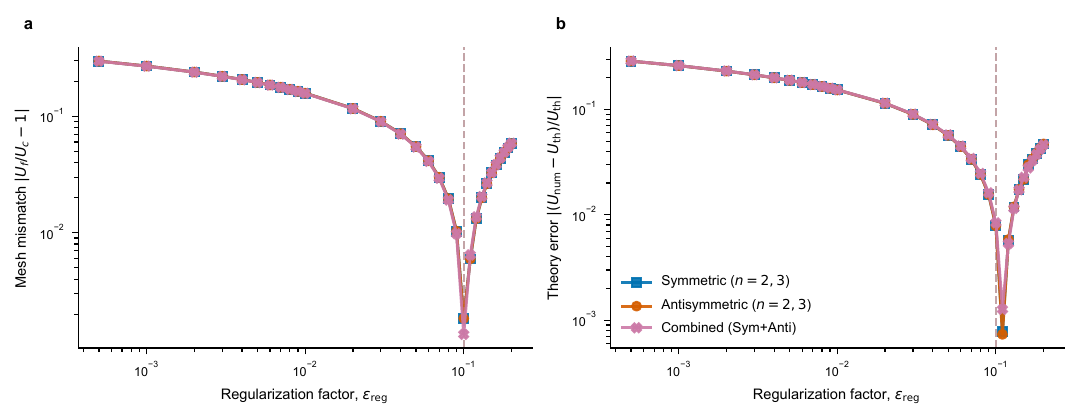}
\caption{\textbf{Regularization parameter selection for the boundary integral solver.}
Both panels show results for three stroke classes (symmetric $n{=}2$, anti-symmetric, and combined), evaluated at deformation amplitude $\varepsilon=0.01$ with coarse resolution $M_c=36$ and fine resolution $M_f=72$ boundary vertices and $T=256$ time steps per period.
(a)~Mesh-convergence indicator $|U_f/U_c - 1|$, measuring the fractional change in computed speed when doubling the spatial resolution from $M_c$ to $M_f$, as a function of the Stokeslet regularization factor $\varepsilon_{\mathrm{reg}}$.
(b)~Theory error $|(U_{\mathrm{num}}-U_{\mathrm{th}})/U_{\mathrm{th}}|$, measuring the deviation of the fine-grid numerical result from the $O(\varepsilon^2)$ analytical prediction.
Dashed vertical lines indicate the value of $\varepsilon_{\mathrm{reg}}$ that minimises the mesh-convergence indicator for each stroke class; this criterion was adopted to determine the per-stroke regularization parameters throughout the paper.}
\label{figS2}
\end{figure*}

\begin{figure*}[h!]
\centering
\includegraphics[width=1.0\linewidth]{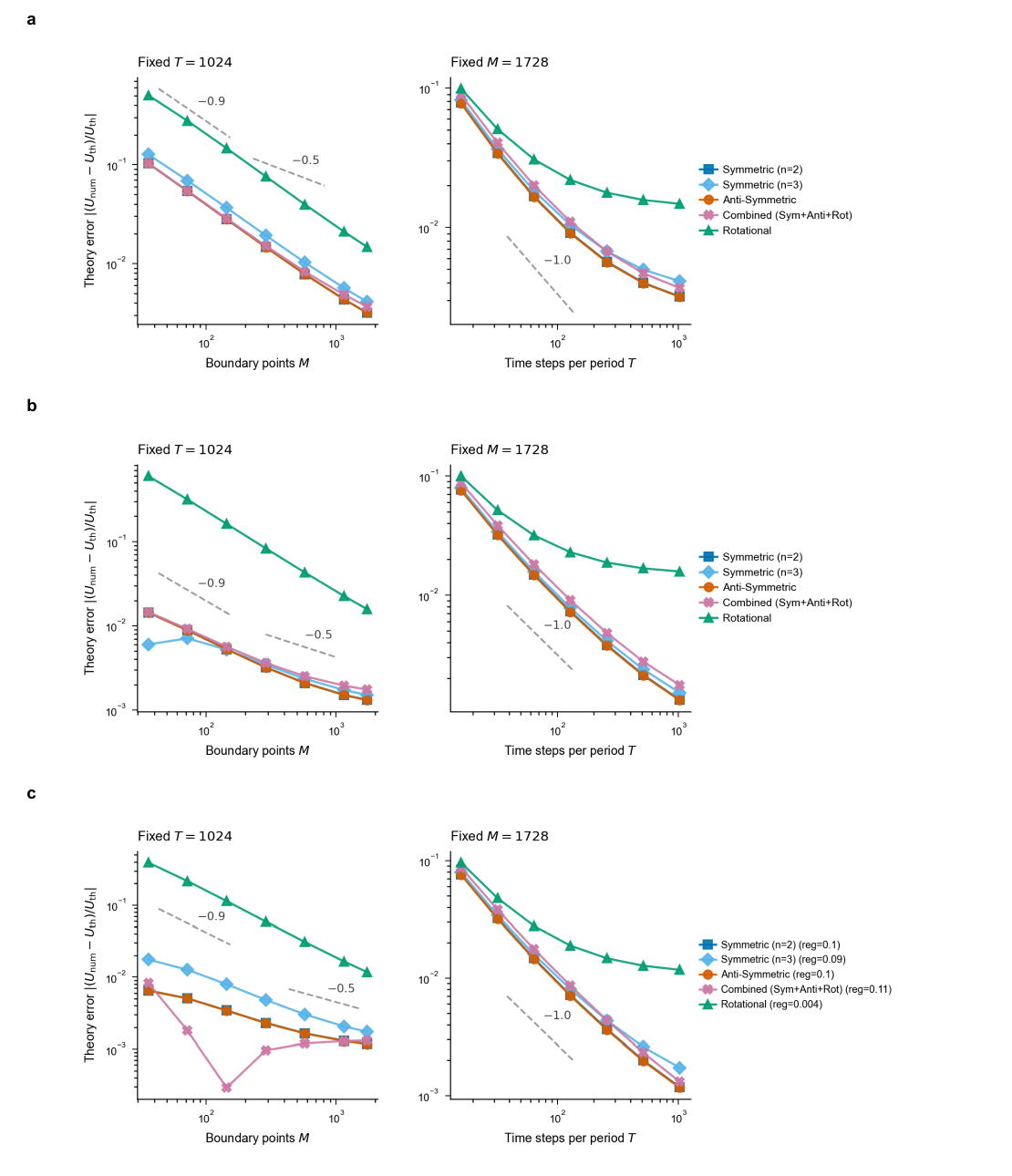}
\caption{\textbf{Convergence of the regularized boundary integral method to small-amplitude theory.}
Theory errors $|(U_\mathrm{num}-U_\mathrm{th})/U_\mathrm{th}|$ for the five stroke classes, shown as a function of spatial resolution (number of boundary vertices $M$; left column) and temporal resolution (number of time steps per period $T$; right column).
(a) Fixed regularization $\varepsilon_{\mathrm{reg}}=0.05$ and fixed deformation amplitude $\varepsilon=0.01$.
(b) Fixed regularization $\varepsilon_{\mathrm{reg}}=0.095$ and fixed deformation amplitude $\varepsilon=0.01$.
(c) Per-stroke optimized regularization $\varepsilon_{\mathrm{reg}}$ and fixed deformation amplitude $\varepsilon=0.01$, illustrating systematic reduction of discretization-induced discrepancies.}
\label{figS3}
\end{figure*}

\begin{figure*}[h!]
\centering
\includegraphics[width=1.0\linewidth]{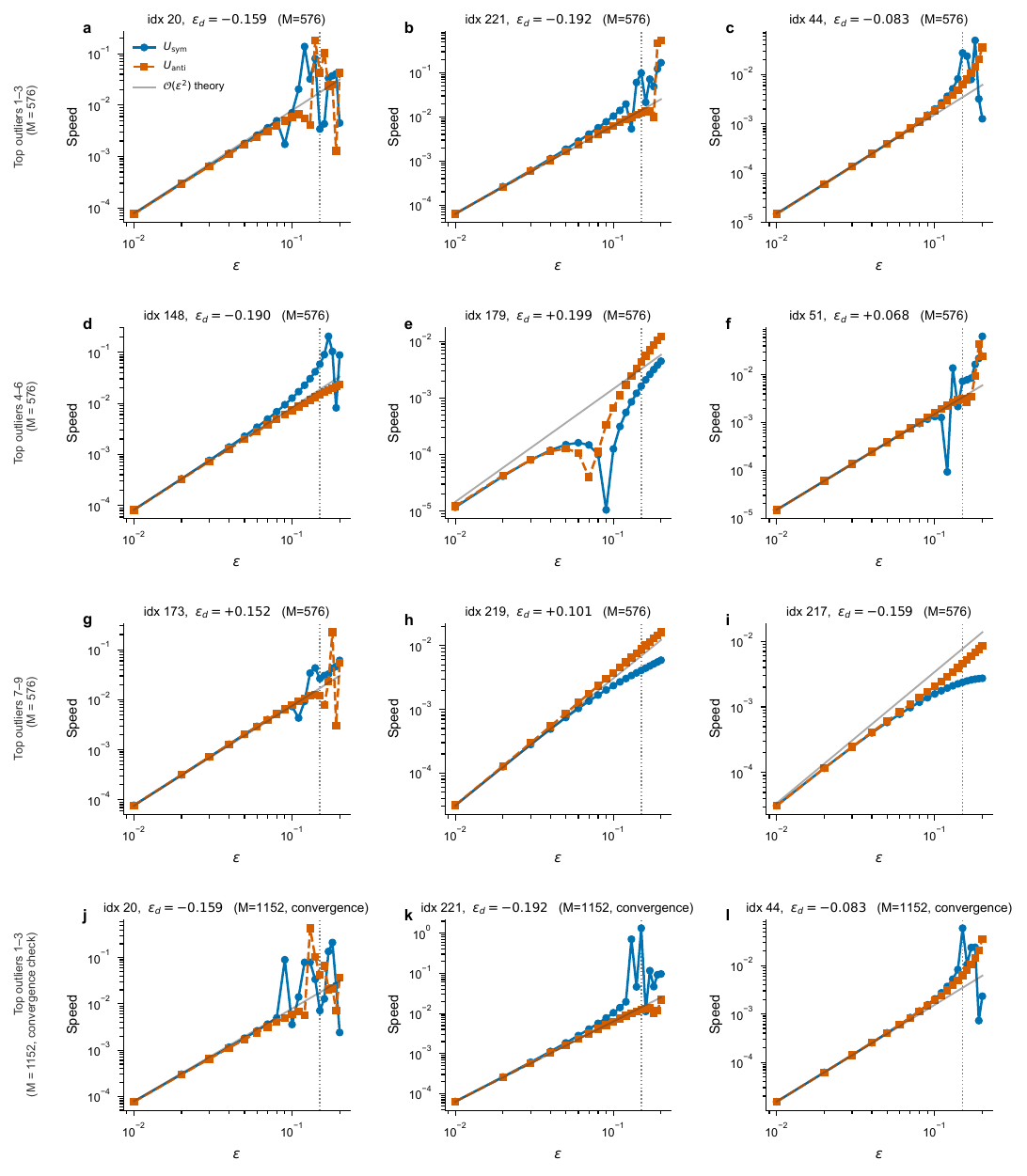}
\caption{\textbf{Epsilon sweep for duality outlier swimmers.}
Panels a--i: Swimming speed of the symmetric stroke (blue) and its anti-symmetric dual (orange) as a function of deformation amplitude $\varepsilon$ for the nine strongest duality-breaking outliers identified in Fig.~\ref{fig3}b, simulated at $M=576$ boundary vertices. The dotted vertical line marks the nominal value $\varepsilon=0.15$ used in Fig.~\ref{fig3}b. At small $\varepsilon$, both strokes closely follow the $O(\varepsilon^2)$ analytical prediction (grey) and their speeds agree, confirming that duality holds in the perturbative regime. Deviations at larger $\varepsilon$ arise through two mechanisms: near-cancellation of the leading-order coefficient (causing the measured speed to be dominated by higher-order contributions; panels a--e) or irregular divergence of the numerical velocity from the perturbative prediction as the deformation amplitude grows (panels f--i). Panels j--l: The top three outliers re-simulated at $M=1152$ vertices (convergence check), showing that the duality-breaking behaviour persists under mesh refinement and is not a numerical artefact.
}
\label{figS4}
\end{figure*}

\clearpage
\newpage

\subsection*{Supplementary Movies}

\noindent\textbf{Supplementary Movie~1.}
Swimming motion of eight two-dimensional deformable swimmers at deformation amplitude $\varepsilon = 0.15$. The top row shows four symmetric-class swimmers using radial ($\alpha_n$) and tangential ($\beta_n$) modes with wavenumbers $n = 2{,}3$ (two variants), $n = 3{,}4$, and a generic multi-mode stroke; the bottom row shows corresponding anti-symmetric swimmers using $\gamma_n$ and $\delta_n$ modes (two variants of $n = 2{,}3$, one of $n = 3{,}4$), and a rotational swimmer ($\alpha_3$, $\gamma_3$). Coloured dots on the boundary track material points, and the trailing line marks the centre-of-mass trajectory. Symmetric and anti-symmetric swimmers in each column achieve comparable net displacement, illustrating hydrodynamic duality.

\noindent\textbf{Supplementary Movie~2.}
Swimming motion of the two optimal-efficiency swimmers identified in Fig.~\ref{fig2}c, at deformation amplitude $\varepsilon = 0.15$. Left: optimal stroke using Fourier modes $n = 2$ and $3$; right: optimal stroke using modes $n = 3$ and $4$. The deformation amplitudes correspond to the eigenvectors of the efficiency maximisation problem (Eqs.~\eqref{eq:eta_star} and \eqref{eq:eta_star_n3}).

\noindent\textbf{Supplementary Movie~3.}
Swimming motion of the three-dimensional prolate ellipsoid swimmer ($L_z/L_x = 1.5$, $\varepsilon = 0.15$) with symmetric $n = 2{,}3$ deformation modes, corresponding to the schematic in Fig.~\ref{fig3}c. Left panel: deforming surface mesh. Right panel: animated cross-sections of the body at several heights along the dorsal--ventral axis, showing the planar Fourier-mode shape changes applied at each latitude.

\clearpage
\newpage

\bibliography{swimmerBib}

\end{document}